\documentclass[3p,a4paper]{elsarticle}
\usepackage[utf8]{inputenc}
\usepackage{amssymb}
\usepackage{gensymb}
\usepackage{booktabs}
\usepackage[normalem]{ulem}
\usepackage{listings}
\usepackage{siunitx}
\usepackage[hidelinks]{hyperref}
\hypersetup{
    colorlinks,
    linkcolor={red!50!black},
    citecolor={blue!50!black},
    urlcolor={blue!80!black}
}
\usepackage{caption}
\usepackage{subcaption}
\usepackage{lineno}

\usepackage[usenames,dvipsnames]{xcolor}
\definecolor{TODOcolor}{HTML}{888800}

\definecolor{FRESHaColor}{HTML}{000088} 

\definecolor{revAcolor}{HTML}{000000}
\newcommand{\revA}[1]{\textcolor{revAcolor}{#1}}

\definecolor{revBcolor}{HTML}{000000}
\newcommand{\revB}[1]{\textcolor{revBcolor}{#1}}

\newcommand{\Tab}[1]{Tab.~\ref{tbl:#1}}
\newcommand{\Fig}[1]{Fig.~\ref{fig:#1}}

\begin{document}
\newcommand{\fun}[1]{\texttt{#1}}
\newcommand{\cfun}[1]{\texttt{#1} (complex)} 
\newcommand{\ofun}[1]{\texttt{#1}} 
\newcommand{\tReal}{\texttt{Real}}
\newcommand{\tfloat}{\texttt{float}}
\newcommand{\tdouble}{\texttt{double}}
\newcommand{\tldouble}{\texttt{long double}}
\newcommand{\tfloatHTE}{\texttt{float128}}
\newcommand{\tMPFR}{\texttt{mpfr}}
\newcommand{\tBBFLOAT}{\texttt{cpp\_bin\_float}}
\newcommand{\colCentr}[1]{\multicolumn{1}{c}{#1}}
\newcommand{\MR}[1]{\href{https://gitlab.com/yade-dev/trunk/-/merge_requests/#1}{!#1}}
\newcommand{\ISSUE}[1]{\href{https://gitlab.com/yade-dev/trunk/-/issues/#1}{\##1}}

\begin{frontmatter}
    \title{Implementation of high-precision computation capabilities into the open-source dynamic simulation framework YADE}

    \author[gda1,gda2]{Janek Kozicki}
    \address[gda1]{Faculty of Applied Physics and Mathematics, Gdańsk University of Technology, 80-233 Gdańsk, Poland}
    \address[gda2]{Advanced Materials Center, Gdańsk University of Technology, 80-233 Gdańsk, Poland}
    \ead{Corresponding author: jkozicki@pg.edu.pl, ORCID: 0000-0002-8427-7263}

    \author[fre1]{Anton Gladky}
    \address[fre1]{Institute for Mineral Processing Machines and Recycling Systems Technology, TU Bergakademie Freiberg, 09599 Freiberg, Germany}
    \ead{Anton.Gladky@iart.tu-freiberg.de, ORCID: 0000-0001-5270-9060}

    \author[newc1]{Klaus Thoeni}
    \address[newc1]{Centre for Geotechnical Science and Engineering, The University of Newcastle, 2308 Callaghan, Australia}
    \ead{klaus.thoeni@newcastle.edu.au, ORCID: 0000-0001-7351-7447}

    \date{\today}

    \journal{Computer Physics Communications}
    \begin{abstract}
        This paper deals with the implementation of arbitrary precision calculations into the open-source discrete element framework
        YADE published under the GPL-2+ free software license.
        \revA{This new capability paves the way for the simulation framework to be used in many new fields such as quantum mechanics.}
        The implementation details and associated gains in the accuracy of the results are discussed. Besides the ,,standard'' \texttt{double} (64 bits) type, support for the
        following high-precision types is added: \texttt{long double} (80 bits), \texttt{float128} (128 bits),
        \texttt{mpfr\_float\_backend} (arbitrary precision) and \texttt{cpp\_bin\_float} (arbitrary precision).
        Benchmarks are performed to quantify the additional computational cost involved with the new supported precisions.
        Finally, a simple calculation of a chaotic triple pendulum is performed to demonstrate the new capabilities and the effect of different precisions on the simulation result.
    \end{abstract}

    \begin{keyword}
        arbitrary accuracy
        \sep multiple precision arithmetic
        \sep dynamical systems \texttt{05.45.-a}
        \sep computational techniques \texttt{45.10.-b}
        \sep numerical analysis \texttt{02.60.Lj}
        \sep error theory \texttt{06.20.Dk}
    \end{keyword}
\end{frontmatter}


\section{Introduction}

The advent of a new era of scientific computing has been predicted in the literature, one in which the numerical precision required for
computation is as important as the algorithms and data structures in the program~\cite{Bailey2012,Kahan2004,Kahan2006,McCurdy2002}. An appealing example of a simple
computation gone wrong was presented in the talk ``Are we just getting wrong answers faster?'' of Stadtherr in 1998~\cite{Stadtherr1998}.
An exhaustive list of such computations along with a very detailed analysis can be found in~\cite{Kahan2006}.

Many examples exist where low-precision calculations resulted in disasters. The military identified an accumulated error in multiplication by a constant factor of $0.1$, which has no exact
binary representation, as the cause for a Patriot missile failure
on 25 February 1991, which resulted in several fatalities~\cite{Blair1992}.
\revB{If more bits were used to represent a number,} the explosion of an Ariane 5 rocket
launched by the European Space Agency on 4 June 1996 could have been prevented~\cite{Lions1996,Lann2006,Jezequel1997,BenAri2001} as it was a
result of \revB{an inappropriate conversion from a 64 bit floating point number into a 16 bit signed integer. Indeed, the 64 bit floating point number was too big to be represented as a 16 bit signed integer}.
\revB{On 14 May 1992 the rendezvous between the shuttle Endeavour and the Intelsat~603 spacecraft nearly failed. The problem was traced back to a mismatch in precision~\cite{Peterson1996,Huckle2019}.
More catastrophic failures related to the lack of precision are discussed in~\cite{Peterson1996,Huckle2019}}.
In 2012 it was predicted that most future technical computing
will be performed by people with only basic training in numerical analysis \revB{or none at all~\cite{Loh2002,Goldberg1991,Bailey2012,Kahan2006}}.
High-precision computation is an attractive option for such
users, because even if a numerically better algorithm \revB{with smaller error or faster convergence is known for a given problem (e.g.~Kahan summation~\cite{Goldberg1991} for avoiding accumulating
errors\footnote{\label{precision_benefits}for $n$ summands and $\varepsilon$ Unit in Last Place (ULP) error, the error in regular summation is $n\varepsilon$,
error in Kahan summation is $2\varepsilon$, while error with regular summation in twice higher precision is $n\varepsilon^2$. See proof of Theorem 8~in~\cite{Goldberg1991}.})},
it is often easier and more efficient to increase the
precision of an existing algorithm rather \revB{than deriving}
\revB{and} implementing a new one~\cite{Bailey2012,Kahan2004} --- a feat which is made possible by the work presented
in this paper.
It shall however be noted that increasing precision is not the answer to all types of problems, as recently a new kind of
a pathological systematic error of up to $14\%$ has been discovered in a certain type of Bernoulli map calculations which cannot be mitigated by increasing
the precision of the calculations~\cite{Boghosian2019}.
\revB{In addition, switching to high-precision generally means longer run times~\cite{Isupov2020,Fousse2007}}.

Nowadays, high-precision calculations find application in various different
domains, such as
long-term stability analysis of the solar system~\cite{Laskar2009,Sussman1992,Bailey2012}, supernova simulations~\cite{Hauschildt1999}, climate modeling~\cite{He2001}, Coulomb n-body atomic
simulations~\cite{Bailey2002,Frolov2004}, studies of the fine structure constant~\cite{Yan2003,Zhang1996},
\revA{identification of constants in quantum field theory~\cite{Broadhurst1999,Bailey2005a}}, numerical integration in
experimental mathematics~\cite{Bailey2005b,Lu2010}, \revA{three-dimensional incompressible Euler flows~\cite{Caflisch1993}, fluid undergoing vortex sheet roll-up~\cite{Bailey2005a}}, integer relation detection~\cite{Bailey2000}, finding sinks in the Henon Map~\cite{Joldes2014}
and iterating the Lorenz attractor~\cite{Abad2011}. There are many more yet unsolved \revB{high-precision} problems~\cite{Stefanski2013},
\revA{especially in quantum mechanics and quantum field theory where calculations are done with 32, 230 or even 10000 decimal digits of precision~\cite{Pachucki2005,Silkowski2020,Broadhurst1999}}.
\revA{Additionally Debian, a Linux distribution with one of the largest archive of packaged free software is now moving numerous numerical computation packages
such as Open MPI, PETSc, MUMPS, SuiteSparse, ScaLAPACK, METIS, HYPRE, SuperLU, ARPACK and others into 64-bit builds~\cite{Debian}.
In order to stay ahead of these efforts, simulations frameworks need to pave the way into 128-bit builds and higher.}

The open-source dynamic simulation framework YADE~\cite{Kozicki2008,yade:doc2} is extensively used by many researchers all over the world with a large, active and growing community of
more than 25 contributors.
\revA{YADE, which stands for ``Yet Another Dynamic Engine'', was initially developed as a generic dynamic simulation framework.
The computation parts are written in C++ using flexible object models, allowing independent implementation of new algorithms
and interfaces. Python (interpreted programming language, which wraps most of C++ YADE code) is used for rapid and concise scene construction, simulation control, postprocessing and debugging.
Over the last decade YADE has evolved into a powerful discrete element modelling package.}
The framework benefits from a great amount of features added by the YADE community, for example particle fluid
coupling~\cite{Gladky2014,Lomine2011,Maurin2015}, thermo–hydro-mechanical coupling~\cite{Caulk2020,Krzaczek2019,Krzaczek2020},
interaction with deformable membrane-like structures, cylinders and grids~\cite{Effeindzourou2016,Bourrier2013,Thoeni2013},
FEM-coupling~\cite{Jerier2011,Frenning2008,Guo2014}, polyhedral particles~\cite{Boon2013,Elias2014,Gladky2017}, deformable
particles~\cite{Haustein2017},
brittle materials~\cite{ScholtesDonze2013,Donze2021},
quantum dynamics of diatomic molecules~\cite{Jasik2018,Jasik2021}
and many others. A more extensive list of publications involving the use of YADE can be found on the framework's web page~\cite{YadePubs}.
A list of selected available YADE modules and features is presented in~\Tab{supported_packages}.
\revA{Although its current focus is on discrete element simulations of granular material, its modular design allows it to be easily extended to new applications that rely on high-precision calculations.}

The present work deals with the implementation of high-precision support for YADE \revA{which will open the way for YADE to be used in many new research areas such as
quantum mechanics~\cite{Jasik2018,Jasik2021}, special relativity, general relativity, cosmology, quantum field theory
and conformal quantum geometrodynamics~\cite{Santamato2015,DeMartini2013}}. The programming techniques necessary for such extension are
presented and discussed in Section~\ref{sec:Implementation}. Relevant tests and speed benchmarks are performed in
Sections~\ref{sec:Testing} and~\ref{sec:Benchmark}. A simple chaotic triple pendulum simulation with high precision is presented
in Section~\ref{sec:Simulation}.
Finally, conclusions are drawn and it is discussed how this new addition to the framework will enable research in many new directions.

\begin{table*}[ht]
\caption{Selected modules and features in YADE.}
\label{tbl:supported_packages}
\begin{center}
\begin{tabular}{l*{2}{l}}
\addlinespace[3pt]
\toprule
cmake flag						& description \\
\midrule
							& \multicolumn{1}{c}{High-precision support in present YADE version\textsuperscript{\ref{supported_modules}}.} \\
\cmidrule(lr){2-2}

(always on)						& Discrete Element Method~\cite{yade:doc2,Kozicki2008}.             \\
(always on)						& Deformable structures~\cite{Effeindzourou2016,Bourrier2013,Thoeni2013}. \\
\footnotesize{\texttt{ENABLE\_CGAL}}			& Polyhedral particles, polyhedral particle breakage~\cite{Elias2014,Gladky2017}. \\
\footnotesize{\texttt{ENABLE\_LBMFLOW}}			& Fluid-solid interaction in granular media with coupled \\
							& Lattice Boltzmann/Discrete Element Method~\cite{Lomine2011}. \\
\footnotesize{\texttt{ENABLE\_POTENTIAL\_PARTICLES}}	& Arbitrarily shaped convex particle described as a 2nd degree \\
							& polynomial potential function~\cite{Boon2013}. \\
\midrule
							& \multicolumn{1}{c}{Selected YADE features with high-precision support.} \\
\cmidrule(lr){2-2}

\footnotesize{\texttt{ENABLE\_VTK}}			& Exporting data and simulation geometry to ParaView~\cite{yade:doc2} \\
(always on)						& Importing geometry from CAD/CAM software (\texttt{yade.ymport})~\cite{yade:doc2}. \\
\footnotesize{\texttt{ENABLE\_ASAN}}			& AddressSanitizer allows detection of memory errors, memory leaks, \\
							& heap corruption errors and out-of-bounds accesses~\cite{asan}. \\
\footnotesize{\texttt{ENABLE\_OPENMP}}			& OpenMP threads parallelization, full support for \tdouble, \\
							& \tldouble, \tfloatHTE{} types\textsuperscript{\ref{mparray_problem}}.\\
\midrule
							& \multicolumn{1}{c}{Modules under development for high-precision support.} \\
\cmidrule(lr){2-2}
\footnotesize{\texttt{ENABLE\_MPI}}			& MPI environment for massively parallel computation~\cite{yade:doc2}. \\
\footnotesize{\texttt{ENABLE\_VPN}}			& Thermo-hydro-mechanical coupling using virtual pore network~\cite{Krzaczek2019,Krzaczek2020}. \\
\footnotesize{\texttt{ENABLE\_NRQM}}			& Quantum dynamics simulations of diatomic molecules including\\
							& photoinduced transitions between the coupled states~\cite{Jasik2018,Jasik2021}. \\
\bottomrule

\end{tabular}
\end{center}
\end{table*}

\section{Implementation of arbitrary precision}
\label{sec:Implementation}

\subsection{General overview}

Since the beginning of YADE~\cite{Kozicki2008}, the declaration `\texttt{using Real=double;}'\footnote{originally YADE was written in C++03, hence, before the switch to C++17 it was `\texttt{typedef double Real;}'
}
was used as the main floating point type with the intention to use it instead of a plain \texttt{double}
everywhere in the code. The goal of using \tReal{} was to allow replacing its definition with other possible precisions\footnote{see for
    example: \url{https://answers.launchpad.net/yade/+question/233320}}. Hence, the same strategy was followed for other types used in the calculations, such as vectors and matrices. Per definition the
last letter in the type name indicates its underlying type, e.g. '\texttt{Vector3r v;}' is a 3D vector
$\vec{v}\in\widetilde{\mathbb{Q}}^3\in\mathbb{R}^3$, and \texttt{Vector2i} is a 2D vector of integers (where $\widetilde{\mathbb{Q}}$ is a subset of rational
numbers $\mathbb{Q}$, which are representable by the currently used precision: $\widetilde{\mathbb{Q}}\in\mathbb{Q}\in\mathbb{R}$; the name \tReal{} is used instead of {\texttt{Rational}}
or {\texttt{FloatingPoint}} for the sake of brevity).

\newcommand{\libBox}[9]{\put(#1,#2){\framebox(#3,#4){\hbox{\hsize=#5\vtop{\scriptsize#9}}}}\put(#1,#6){\framebox(#7,1.2){\hbox{\hsize=16mm\vtop{\footnotesize#8}}}}}
\newcommand{\libBoxL}[9]{\put(#1,#2){\colorbox{orange}{\framebox(#3,#4){\hbox{\hsize=#5\vtop{\scriptsize#9}}}}}\put(#1,#6){\colorbox{orange}{\framebox(#7,1.2){\hbox{\hsize=16mm\vtop{\footnotesize#8}}}}}}
\newcommand{\libBoxW}[9]{\put(#1,#2){\colorbox{green}{\framebox(#3,#4){\hbox{\hsize=#5\vtop{\scriptsize#9}}}}}\put(#1,#6){\colorbox{green}{\framebox(#7,1.4){\hbox{\hsize=27mm\vtop{\footnotesize#8}}}}}}
\newcommand{\libBoxG}[9]{
\put(#1,#2){\multiput(0,0)(0.872,0){9}{\line(1,0){0.414}}}%
\put(#1,#2){\multiput(0,#4)(0.872,0){9}{\line(1,0){0.414}}}%
\put(#1,#2){\multiput(0,0)(0,0.895){5}{\line(0,1){0.414}}}%
\put(#1,#2){\multiput(#3,0)(0,0.895){5}{\line(0,1){0.414}}}%
\put(#1,#6){\multiput(0,-0.05)(0,0.895){2}{\line(0,1){0.414}}}%
\put(#1,#6){\multiput(6.5,-0.05)(0,0.895){2}{\line(0,1){0.414}}}%
\put(#1,#6){\put(0,1.24){\multiput(0,0)(0.872,0){8}{\line(1,0){0.414}}}}%
\put(#1,#2){\makebox(#3,#4){\hbox{\hsize=#5\vtop{\scriptsize#9}}}}%
\put(#1,#6){\makebox(#7,1.2){\hbox{\hsize=16mm\vtop{\footnotesize#8}}}}%
}

\begin{figure}[t] 
	\setlength{\unitlength}{2.6mm}
	\begin{picture}(64,27)
		\renewcommand{\baselinestretch}{1.1}
		\setlength\fboxsep{0pt}
		\libBoxL{0}{0}{7.4}{4}{18.2mm}{4.05}{6.6}{Boost}{Collection of\\C++ libraries}
		\libBoxL{8}{0}{7.4}{4}{18.2mm}{4.05}{6.6}{CGAL}{Computational\\Geometry\\Algorithms}
		\libBoxL{16}{0}{7.4}{4}{18.2mm}{4.05}{6.6}{OpenGL}{Open\\Graphics\\Library}
		\libBoxL{24}{0}{7.4}{4}{18.2mm}{4.05}{6.6}{VTK}{The\\Visualization\\Toolkit}
		\libBoxL{32}{0}{7.4}{4}{18.2mm}{4.05}{6.6}{LAPACK}{Linear\\Algebra\\Package}
		\libBoxL{40}{0}{7.4}{4}{18.2mm}{4.05}{6.6}{CHOLMOD}{Cholesky\\factorization\\Package}
		\libBoxL{48}{0}{7.4}{4}{18.2mm}{4.05}{6.6}{Other}{Multiple\\different\\C++ libraries}
		\libBoxL{56}{0}{7.4}{4}{18.2mm}{4.05}{6.6}{EIGEN}{Mathematical\\Header-only\\Library}
		\put(3.7,7){\line(1,0){56}}
		\put(3.7,7){\vector(0,-1){1.75}}
		\put(11.7,7){\vector(0,-1){1.75}}
		\put(19.7,7){\vector(0,-1){1.75}}
		\put(27.7,7){\vector(0,-1){1.75}}
		\put(35.7,7){\vector(0,-1){1.75}}
		\put(43.7,7){\vector(0,-1){1.75}}
		\put(51.7,7){\vector(0,-1){1.75}}
		\put(59.7,7){\vector(0,-1){1.75}}
		\libBoxW{11.7}{10}{12}{4}{29.5mm}{14.05}{11}{YADE~C++}{Yet Another\\Dynamic Engine\\(C++ core)}
		\libBoxW{25.7}{10}{12}{4}{29.5mm}{14.05}{11}{YADE~Python}{Yet Another\\Dynamic Engine\\(C++ Python wrapper)}
		\libBoxW{39.7}{10}{12}{4}{29.5mm}{14.05}{11}{MiniEigen(HP)}{Python wrapper\\for some EIGEN-calls}
		\put(17.7,7){\line(0,1){3}}
		\put(25.7,12){\vector(-1,0){2}}
		\put(37.7,12){\vector(1,0){2}}
		\put(45.7,8.5){\line(0,1){1.5}}
		\put(45.7,8.5){\line(1,0){15.3}}
		\put(61,8.5){\vector(0,-1){3.25}}
		\libBox{0}{20}{7.4}{4}{18.2mm}{24.05}{6.6}{DEM}{Discrete\\Element\\Method}
		\libBox{8}{20}{7.4}{4}{18.2mm}{24.05}{6.6}{Grid/PFacet}{Deformable\\structures}
		\libBox{16}{20}{7.4}{4}{18.2mm}{24.05}{6.6}{Polyhedra}{Polyhedral\\particles}
		\libBox{24}{20}{7.4}{4}{18.2mm}{24.05}{6.6}{LBM}{Lattice\\Boltzmann/\\DEM}
		\libBox{32}{20}{7.4}{4}{18.2mm}{24.05}{6.6}{Potential}{Potential\\particles}
		\libBox{40}{20}{7.4}{4}{18.2mm}{24.05}{6.6}{Other}{Multiple\\different\\modules}
		\libBoxG{48}{20}{7.4}{4}{18.2mm}{24.05}{6.6}{{VPN}}{Thermo hydro\\mechanical\\coupling}
		\libBoxG{56}{20}{7.4}{4}{18.2mm}{24.05}{6.6}{{NRQM}}{Nonrelativistic\\Quantum\\Mechanics}
		\put(3.7,17.7){\line(1,0){56}}
		\put(17.7,17.7){\vector(0,-1){2.25}}
		\put(3.7,20){\line(0,-1){2.3}}
		\put(11.7,20){\line(0,-1){2.3}}
		\put(19.7,20){\line(0,-1){2.3}}
		\put(27.7,20){\line(0,-1){2.3}}
		\put(35.7,20){\line(0,-1){2.3}}
		\put(43.7,20){\line(0,-1){2.3}}
		\put(51.7,20){\line(0,-1){2.3}}
		\put(59.7,20){\line(0,-1){2.3}}
	\end{picture}
	\caption{Simplified dependency tree of the open-source framework YADE.
		External dependencies are marked in orange.
		The green boxes indicate parts of the framework that needed to be adapted for high-precision.
		Selected YADE modules which support high-precision are in the top row, dashed lines indicate modules under development (also see~\Tab{supported_packages}).
	}
	\label{fig:dependencies}
\end{figure}
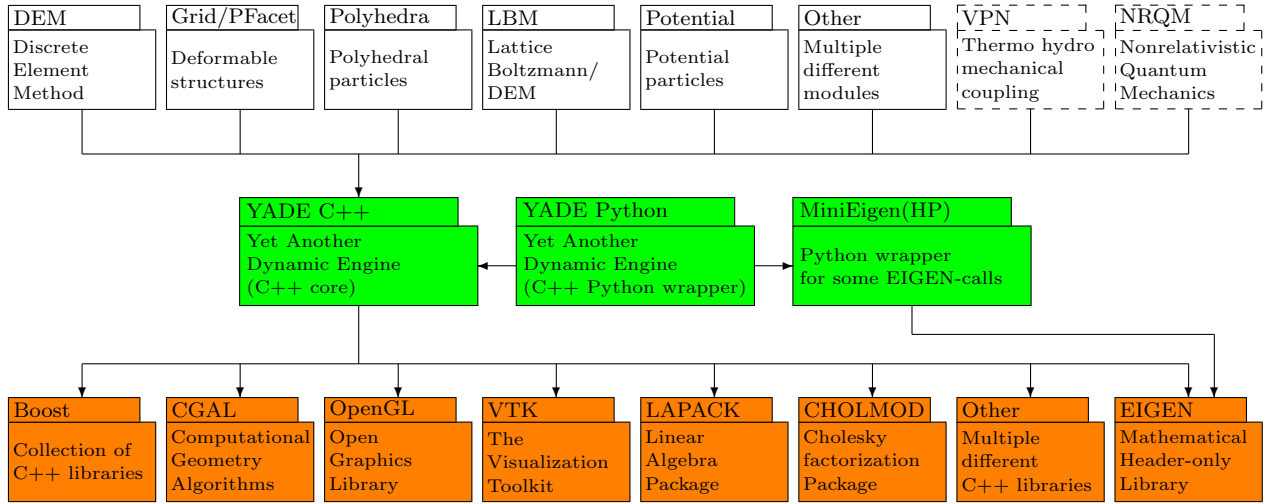

In the presented work, the goal to use high precision is achieved by using the C++ operator overloading functionality
and the \texttt{boost::multiprecision} library. A simplified dependency diagram of YADE
is shown in \Fig{dependencies}. The layered structure of YADE remains nearly the same as in the original paper by Kozicki and Donz\'{e}~\cite{Kozicki2008}. It is built on top
of several well established libraries (marked with orange in \Fig{dependencies}) as discussed in Section~\ref{sec:libraryCompat}.
Some changes were necessary in the structure of the framework (marked with green in \Fig{dependencies}) as highlighted in Section~\ref{sec:backwardCompat}.
The top row in \Fig{dependencies} indicates selected YADE modules with respective citations listed in \Tab{supported_packages}.
\revA{It should be noted that YADE relies on many external libraries to expand its functionality which can result in a demanding server setup.}

The Boost library~\cite{Boost2020} provides convenient wrappers for other high-precision
types with the perspective of adding more such types in the future\footnote{\label{libqd_footnote}at the time of writing, the
quad--double library with 62 decimal places (package \texttt{libqd-dev}) is in preparation, see:
\url{https://github.com/boostorg/multiprecision/issues/184}}. The new supported \tReal{} types are listed in \Tab{supported_types}.
A particular \tReal{} type can be selected during compilation of the code by providing a \texttt{cmake} argument either \texttt{REAL\_PRECISION\_BITS}
or \texttt{REAL\_DECIMAL\_PLACES}\footnote{\label{installation}see \url{http://yade-dem.org/doc/HighPrecisionReal.html} for detailed documentation}.

The process of adding high-precision support to YADE was divided into several stages which are described in the subsections below\footnote{also see the consolidated merge request: \MR{383}}.

\begin{table*}[ht]
\caption{List of high-precision types supported by YADE.}
\label{tbl:supported_types}
\begin{center}
\begin{tabular}{*{6}{l}}
\toprule
~                                      & Total    & Decimal          & Exponent       & Significant       & ~                            \\
Type                                   & bits     & places           & bits           & bits              & Notes                        \\
\toprule
\texttt{float}                         & 32       & 6                & 8              & 24                & only for testing             \\
\texttt{double}                        & 64       & 15               & 11             & 53                & hardware accelerated         \\
\texttt{long double}$^\dag$            & 80       & 18               & 15$^\ddag$     & 64                & hardware accelerated         \\
\texttt{boost float128}$^\mathsection$ & 128      & 33               & 15             & 113               & may be hardware accelerated  \\
\texttt{boost mpfr}$^\mathsection$                    & $N$      & $N \log_{10}(2)$ & ---            & ---               & MPFR library as wrapped by Boost        \\
\texttt{boost cpp\_bin\_float}$^\mathsection$         & $N$      & $N \log_{10}(2)$ & ---            & ---               & uses Boost only, but is slower  \\
\bottomrule

\end{tabular}\\
\begin{flushleft}
$^\dag$~\scriptsize{The specifics of \texttt{long double} depend on the particular compiler and hardware; the values in this table correspond to the most common x86 platform and the \texttt{g++} compiler.}\\
$^\ddag$~\scriptsize{All types use 1 bit to store the sign and all types except \texttt{long double} have an implicit first bit=1, hence here the sum $15+64\neq80$.}\\
$^\mathsection$~\scriptsize{The complete C++ type names for the Boost high-precision types are as follows: \texttt{boost::multiprecision::float128},
\texttt{boost::multiprecision::mpfr\_float\_backend} and \texttt{boost::multiprecision::cpp\_bin\_float}}
\end{flushleft}
\end{center}
\end{table*}

\subsection{Preparations}

To fully take advantage of the C++ Argument Dependent Lookup (ADL), the entire YADE codebase was moved into \texttt{namespace
    yade}\footnote{see: \MR{284}}, thus using the C++ standard capabilities to modularize the namespaces for each software package. Similarly,
the libraries used by YADE such as Boost~\cite{Boost2020}, CGAL~\cite{Cgal2020} and EIGEN~\cite{Eigen2020} reside in their respective
\texttt{boost}, \texttt{CGAL} and \texttt{Eigen} namespaces. After this change, all potential naming conflicts between math functions or types
in YADE and these libraries were eliminated.

Before introducing high precision into YADE it was assumed that \tReal{} is actually a Plain Old Data (POD) \tdouble{} type. It was possible to use the
old C-style \ofun{memset}, \ofun{memcpy}, \ofun{memcmp} and \ofun{memmove} functions which used raw-memory access. However, by doing so the
modern C++ structure used by other high-precision types was completely ignored. For example, the MPFR type may reserve memory and inside its structure store a
pointer to it. Trying to set its value to zero by invoking \ofun{memset} (which sets that pointer to \texttt{nullptr}) leads to a memory leak and a subsequent program failure.
In order to make
\tReal{} work with other types, this assumption had to be removed. Hence, \ofun{memset} calls were replaced with \ofun{std::fill} calls, which
when invoked with a POD type reduce to a (possibly faster) version of \ofun{memset} optimized for a particular type in terms of chunk size
used for writing to the memory.
In addition, C++ template specialization mechanisms allow for invoking with a non-POD type which then utilizes the functionality provided by this specific type, such as calling the specific constructors.
All places in the code which used these four raw-memory access
functions were improved to work with the non-POD \tReal{} type\footnote{\label{mparray_problem}see: \MR{381}, with one
    \href{https://gitlab.com/yade-dev/trunk/-/blob/de1f035d30611e7c40ac69b6e947a874732e61be/lib/base/openmp-accu.hpp\#L20}{exception} which is yet to be evaluated and hence \tMPFR{} and \tBBFLOAT{}
    types work with OpenMP, but do not take full advantage of CPU cache size in class \texttt{OpenMPArrayAccumulator}}. For similar reasons one should
not rely on storing an address of the $n^{th}$ component of a \texttt{Vector3r} or \texttt{Vector2r}\footnote{see: \MR{406}}.

Next, all remaining occurrences of \tdouble{} were replaced with \tReal{}\footnote{these changes were divided into several smaller merge
    requests: \MR{326}, \MR{376}, \MR{394}; there were also a couple of changes such as \texttt{MatrixXd} $\rightarrow$ \texttt{MatrixXr} and
    \texttt{Vector3d} $\rightarrow$ \texttt{Vector3r}.} and the high-precision compilation and testing was added to the gitlab Continuous
Integration (CI) testing pipeline, which guarantees that any future attempts to use \tdouble{} type in the code will fail before merging
such changes into the main branch. Next the \tReal{} type was moved from global namespace into \texttt{yade} namespace\footnote{see:
    \MR{364}} to eliminate any potential problems with namespace pollution\footnote{usually such errors manifest themselves as very unrelated
    problems, which are notoriously difficult to debug, e.g. due to the fact that an incorrect type (with the same name) is used; see:
    \ISSUE{57} and \url{https://bugs.launchpad.net/yade/+bug/528509}.}.

\subsection{Library compatibility}
\label{sec:libraryCompat}

In order to be able to properly interface YADE with all other libraries it was important to make sure that mathematical functions (see
\Tab{testing_functions}) are called for the appropriate type. For example, the EIGEN library would have to call the high-precision \fun{sqrt}
function when invoking a \ofun{normalize} function on a \texttt{Vector3r} in order to properly calculate vector length. Several steps were
necessary to achieve this. First, an inline
redirection\footnote{The recommended practice in such cases is to use the Argument Dependent Lookup (ADL) which lets the compiler pick the
best match from all the available candidate functions~\cite{Alexandrescu2001,Meyers2014,Stroustrup2014,Vandevoorde2017}.
No ambiguity is possible, because such situations would always result in a compiler error. This was done by employing the C++ directives
\texttt{using std::\textit{function};} and \texttt{using boost::multiprecision::\textit{function};}
for the respective \texttt{\textit{function}} and then calling the function unqualified (without namespace qualifier) in the
\href{https://gitlab.com/yade-dev/trunk/-/blob/7c8d1b0e6896745c53b71d91d5fb072badc58774/lib/high-precision/MathFunctions.hpp\#L65}{\texttt{MathFunctions.hpp}}
file.}
to these functions was implemented in namespace
\texttt{yade::math} in the file
\href{https://gitlab.com/yade-dev/trunk/-/blob/master/lib/high-precision/MathFunctions.hpp}{\texttt{MathFunctions.hpp}}. Next, all
invocations in YADE to math functions in the \texttt{std} namespace were replaced with calls to these functions in the \texttt{yade::math}
namespace\footnote{see: \MR{380}, \MR{390}, \MR{391}, \MR{392}, \MR{393}, \MR{397}}.
Functions which take only \tReal{} arguments may omit \texttt{math} namespace specifier and use ADL instead.
Also some fixes were done in EIGEN and CGAL\footnote{see: \url{https://gitlab.com/libeigen/eigen/-/issues/1823} and
    \url{https://github.com/CGAL/cgal/issues/4527}}, although they did not affect YADE directly since it was possible to workaround them.

The C++ type traits is a template metaprogramming technique which allows one to customize program behavior (also called polymorphism)
depending on the type used~\cite{Alexandrescu2001,Meyers2014,Stroustrup2014,Vandevoorde2017}. This decision is done by the compiler
(conditional compilation) due to inspecting the types in the compilation stage (this is called static polymorphism).
Advanced C++ libraries provide hooks (numerical traits) to allow library users to inform the library about the used precision type.
The numerical traits were implemented in YADE for the libraries EIGEN and CGAL\footnote{see files
    \href{https://gitlab.com/yade-dev/trunk/-/blob/master/lib/high-precision/EigenNumTraits.hpp}{\texttt{EigenNumTraits.hpp}},
    \href{https://gitlab.com/yade-dev/trunk/-/blob/master/lib/high-precision/CgalNumTraits.hpp}{\texttt{CgalNumTraits.hpp}} and \MR{412}}
as these were the only libraries supporting such a solution at the time of writing this paper.
EIGEN and CGAL are fully compatible and aware of the entire high-precision code infrastructure in YADE.
Similar treatment would be possible for the Coinor~\cite{Nielsen2002,LougeeHeimer2003} library (used by the class
\href{http://yade-dem.org/doc/yade.wrapper.html?highlight=potentialblock#yade.wrapper.PotentialBlock}{\texttt{PotentialBlock}})
if it would provide numerical traits.
Additionally, an OpenGL compatibility layer
has been added by using inline conversion of arguments from \tReal{} to \tdouble{} for the OpenGL functions as OpenGL drawing functions
use \tdouble{}\footnote{see: \MR{412} and file
    \href{https://gitlab.com/yade-dev/trunk/-/blob/master/lib/opengl/OpenGLWrapper.hpp}{\texttt{OpenGLWrapper.hpp}}. If the need for
    drawing on screen with precision higher than \tdouble{} arises (e.g. at high zoom levels) it will be rectified in the future.}.
The VTK compatibility layer was added using a similar approach. Virtual functions were added to convert the \tReal{} arguments to \tdouble{}\footnote{see: \MR{400} and
    \href{https://gitlab.com/yade-dev/trunk/-/blob/master/lib/compatibility/VTKCompatibility.hpp}{\texttt{VTKCompatibility.hpp}}. If the VTK
    display software will start supporting high precision, this solution can be readily improved.} in classes derived from the VTK parent class (such as \texttt{vtkDoubleArray}).

The LAPACK compatibility layer was provided as well, this time highlighting the problems of interfacing with languages which do not support
static polymorphism. The routines in LAPACK are written in a mix of Fortran and C, and have no capability to use high-precision numerical traits like EIGEN
and CGAL. The only way to do this (apart from switching to another library) was to down-convert the arguments to \tdouble{} upon calling
LAPACK routines (e.g.~a routine to solve a linear system) then up-converting the results to \tReal{}. This was the first step to phase out
YADE's dependency on LAPACK. With this approach the legacy code works even when high precision is enabled although the obtained
results are low-precision\footnote{see: \MR{379} and
    \href{https://gitlab.com/yade-dev/trunk/-/blob/master/lib/compatibility/LapackCompatibility.cpp}{\texttt{LapackCompatibility.cpp}}.}.
Additionally, this allows one to test the new high-precision code against the low-precision version when replacing these function calls
with appropriate function calls from another library such as EIGEN in the future. Fortunately, \revB{only two YADE modules depend on LAPACK: potential particles and the flow engine~\cite{Caulk2019}. The latter also depends on CHOLMOD, which also supports the \tdouble{} type only, hence it is not shown in \Tab{supported_packages}. Nevertheless, a similar solution as currently implemented for LAPACK can be used in the future to remove the current dependency on CHOLMOD.}

\subsection{Double, quadruple and higher precisions}
\label{sec:RealHP}

Sometimes a critical section of the computations in C++ would work better if performed in a higher precision\textsuperscript{\ref{precision_benefits}}. This would also guarantee that the overall results in the default precision are correct.
The \texttt{RealHP<N>} types serve this purpose.
In analogy to \tfloat{} and \tdouble{} types used on older systems, the types \texttt{RealHP<2>}, \texttt{RealHP<4>} and \texttt{RealHP<N>} correspond
to double, quadruple and higher multipliers of the \tReal{} precision selected during compilation, e.g. with \texttt{REAL\_DECIMAL\_PLACES}\textsuperscript{\ref{installation}}, respectively.
A simple example where this can be useful is solving a system of linear equations where some coefficients
are almost zero. The old rule of thumb to \textit{,,perform all computation in arithmetic with somewhat more than twice as many
    significant digits as are deemed significant in the data and are desired in the final results''} works well in many cases~\cite{Kahan2004}.
Nevertheless, maintaining a high quality scientific software package without being able to use, when necessary, arithmetic precision
twice as wide can badly inflate costs of development and maintenance~\cite{Kahan2004}.
On the one hand, there might be additional costs for the theoretical formulation of such tricky single-precision problems.
On the other hand, the cost of extra demand for processor cycles and memory when
using \texttt{RealHP<N>} types is picayune when compared with the cost of a numerically adept mathematician's time~\cite{Kahan2006}.
Hence, the new \texttt{RealHP<N>} makes high and multiple-precision simulations more accessible to the researcher community.

The support for higher precision multipliers was added in YADE\footnote{see: \MR{496}} in such a way that \texttt{RealHP<1>}
is the \tReal{} type from~\Tab{supported_types} and every higher number \texttt{N} is a multiplier of the \tReal{} precision.
All other types follow the same naming pattern:
\texttt{Vector3rHP<1>} is the regular \texttt{Vector3r} and \texttt{Vector3rHP<N>} uses the precision multiplier~\texttt{N}. A similar concept is used for
CGAL types (e.g. \texttt{CGALtriangleHP<N>}). One could then use an EIGEN algorithm for solving a system of linear equations with a
higher \texttt{N} using \texttt{MatrixXrHP<N>} to obtain the result with higher precision. Then, after the critical code section, one could potentially continue
the calculations in the default \tReal{} precision.
On the Python side the mathematical functions for the higher precision types are accessible via \texttt{yade.math.HP2.*}. By default only the \texttt{RealHP<2>}
is exported to Python. One can export to Python all the higher types for debugging purposes by adjusting \texttt{\#define YADE\_MINIEIGEN\_HP} in the file
\href{https://gitlab.com/yade-dev/trunk/blob/master/lib/high-precision/RealHPConfig.hpp}{\texttt{RealHPConfig.hpp}}.

On some occasions it is useful to have an intuitive up-conversion between C++ types of
different precisions, say for example to add \texttt{RealHP<1>} to \texttt{RealHP<2>}. The file
\href{https://gitlab.com/yade-dev/trunk/-/blob/master/lib/high-precision/UpconversionOfBasicOperatorsHP.hpp}{\texttt{UpconversionOfBasicOperatorsHP.hpp}}
serves this purpose. After including this header, operations using two different precision types are possible and the resultant
type of such operation will always be the higher precision of the two types.
This header should be used with caution (and only in \texttt{.cpp} files) in order to still be able to take advantage of the C++ static type checking mechanisms.
\revB{As mentioned in the introduction, this type checking whether a number is being converted to a fewer digits representation can prevent mistakes such as the
explosion of the rocket Ariane 5~\cite{Lions1996,Lann2006,Jezequel1997,BenAri2001}.}

\subsection{Backward compatibility with older YADE scripts}
\label{sec:backwardCompat}

In the present work, preserving the backward compatibility with existing older YADE Python scripts was of prime importance. To obtain this the
MiniEigen Python library had to be incorporated into YADE's codebase. The reason for this was the following: \texttt{python3-minieigen} was a binary
package, precompiled using \tdouble{}. Thus any attempt of importing MiniEigen into a YADE Python script (i.e.~using \texttt{from~minieigen~import~*}) when YADE was
using a non-\tdouble{} type resulted in failure. This, combined with the new capability in YADE to use any of the current and future
supported types (see \Tab{supported_types}) would place a requirement on \texttt{python3-minieigen} that it either becomes a header-only
library or is precompiled with all possible high-precision types. It was concluded that integrating its source directly into YADE is the most
reasonable solution. Hence, old YADE scripts that use supported modules\textsuperscript{\ref{supported_modules}} can be immediately converted
to high precision by switching to \texttt{yade.minieigenHP}. In order to do so, the following line:
\begin{lstlisting}
from minieigen import *
\end{lstlisting}
\noindent
has to be replaced with:
\begin{lstlisting}
from yade.minieigenHP import *
\end{lstlisting}
Respectively \texttt{import minieigen} has to be replaced with \texttt{import yade.minieigenHP as minieigen}, the old name \texttt{as
    minieigen} being used for the sake of backward compatibility with the rest of the script.

Python has native support\footnote{see:
    \href{https://gitlab.com/yade-dev/trunk/blob/master/lib/high-precision/ToFromPythonConverter.hpp}{\texttt{ToFromPythonConverter.hpp}} file.}
for high-precision types using the \texttt{mpmath} Python package.
However, it shall be noted that although the coverage of YADE's basic testing and checking (i.e.~\texttt{yade --test} and \texttt{yade --check}) is fairly large, there may still be some
parts of Python code that were not yet migrated to high precision and may not work well with the \texttt{mpmath} module. If such problems
occur in the future, the solution is to put the non compliant Python function into the
\href{https://gitlab.com/yade-dev/trunk/blob/master/py/high-precision/math.py}{\texttt{py/high-precision/math.py}} file\footnote{see also:
    \MR{414}}.

A typical way of ensuring correct treatment of \tReal{} in Python scripts is to initialize Python variables using
\href{http://yade-dem.org/doc/yade.math.html#yade.math.Real}{\texttt{yade.math.Real(arg)}}.
If the initial argument is not an integer and not an \texttt{mpmath} type then it has to be passed as a string (e.g. \texttt{yade.math.Real('9.81')}) to prevent Python from converting it to \tdouble.
Without this special initialization step a mistake can appear in the Python script where the
default Python floating-point type \tdouble{} is for example multiplied or added to the \tReal{} type resulting in a loss of precision\footnote{see: \MR{604} and
    commit \href{https://gitlab.com/yade-dev/trunk/-/commit/494548b82d84ac6467e3a07d3c36e7201247194f}{494548b82d}, where a small change in the Python script
    enabled it to work for high precision.}.

\section{Testing}
\label{sec:Testing}

It should be noted that it is near to impossible to be absolutely certain about the lack of an error in a code~\cite{Kahan2006}.
Therefore, to briefly test the implementation of all mathematical functions available in C++ in all precisions, the following test was
implemented in \href{https://gitlab.com/yade-dev/trunk/-/blob/master/py/high-precision/\_RealHPDiagnostics.cpp}{\texttt{\_RealHPDiagnostics.cpp}}
and run using the Python script \href{https://gitlab.com/yade-dev/trunk/-/blob/master/py/tests/testMath.py}{\texttt{testMath.py}}.
Each available function was evaluated $2\times10^7$ times with on average evenly spaced pseudo-random argument in the range $(-100,100)$.
These $2\times10^7$ evaluations were divided into two sets. The first $10^7$ evaluations were performed with uniformly distributed pseudo-random numbers in the range $(-100,100)$.
The second $10^7$ evaluations were done by randomly displacing a set of $10^7$ equidistant points in the range $(-100,100)$,
where each point was randomly shifted by less than $\pm0.5$ of the distance between the points. This random shift was to lower the
chances of duplicate calculations on the same argument after adjusting to the function domain.
The $2\times10^7$ arguments were subsequently modified to match the domain argument range of each function using simple operations,
such as \fun{abs($\bullet$)} or \fun{fmod(abs($\bullet$),2)-1}.
The obtained result for each evaluation was then compared against its respective \texttt{RealHP<4>} type with four times higher precision.
Care was taken to exactly use the same argument for a higher precision function call. The arguments were randomized and adjusted to the function domain range
in the lower precision \texttt{RealHP<1>}, then the argument was converted to the higher precision by using \texttt{static\_cast}, thereby ensuring that all the extra bits
in the higher precision are set to zero.

The difference expressed in terms of Units in the Last Place (ULP)~\cite{Kahan2006} was calculated. The obtained errors are listed in \Tab{testing_functions}.
During the tests a bug in the implementation of the \fun{tgamma} function for \texttt{boost::multiprecision::float128} was discovered
but it was immediately fixed by the Boost developers\footnote{see:~\url{https://github.com/boostorg/math/issues/307}~}. Some other bug
reports\footnote{see: \url{https://github.com/boostorg/multiprecision/issues/264} and \url{https://github.com/boostorg/multiprecision/issues/262}} instigated a discussion
about possible ways to fix the few problems found with the \texttt{cpp\_bin\_float} type which can be seen in the last column of \Tab{testing_functions}.
A smaller version of this test, with only $2\times10^4$ pseudo-random evaluations\footnote{because this test is time consuming it is not possible to run the
    test involving $2\times10^7$ evaluations in the GitLab CI pipeline after each \texttt{git push}.} was then added to the standard \texttt{yade --test} invocation.

Finally, an AddressSanitizer~\cite{asan,Bellotti2021} was employed to additionally check the correctness of the implementation in the code and to quickly locate memory access bugs. Several critical errors were fixed due to the reports of this sanity checker.
This tool is now integrated into the Continuous Integration (CI) pipeline for the whole YADE project to prevent introduction of such errors
in the future~\cite{Bellotti2021} (\href{https://gitlab.com/yade-dev/trunk/-/blob/57e3849a62135ceb1b12f61ffdfd2cb1065825a3/.gitlab-ci.yml#L243}{\texttt{make\_asan\_HP}} job in the GitLab CI pipeline).

\section{Benchmark}
\label{sec:Benchmark}

\begin{figure}
\begin{center}
	\hbox{\hsize=38mm
		\vtop{\includegraphics[width=38mm]{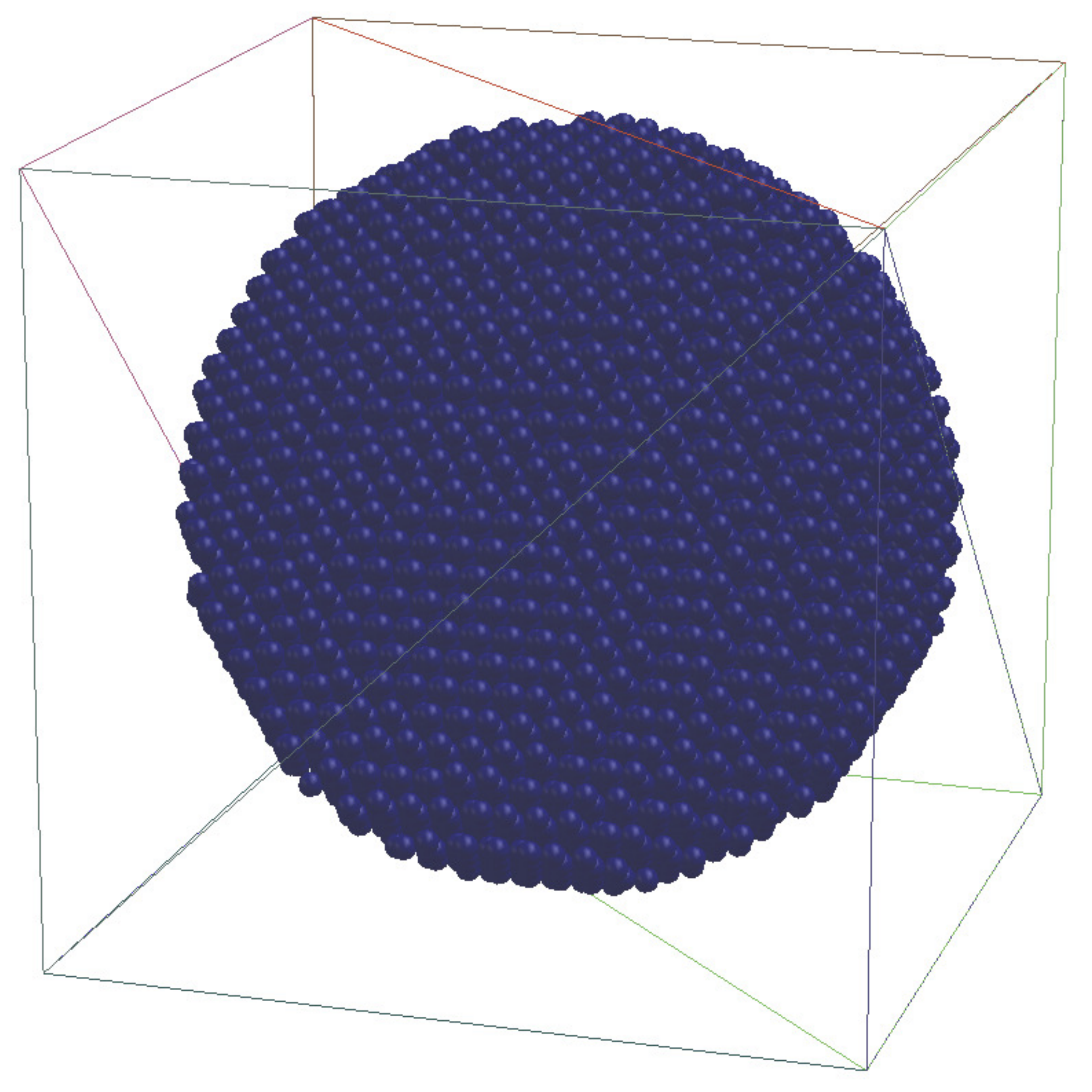}\\(a)}%
		\vtop{\includegraphics[width=38mm]{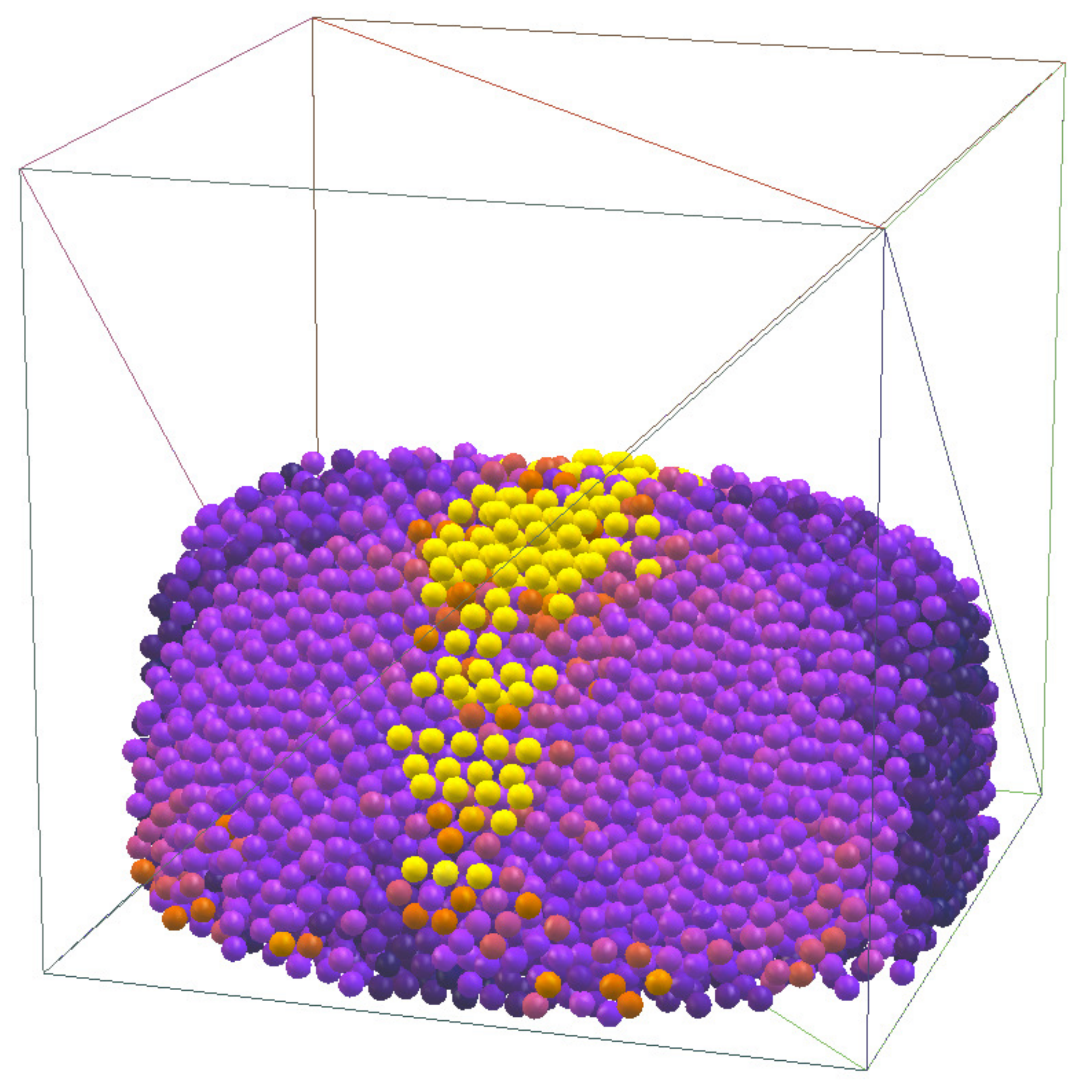}\\(b)}%
		\vtop{\includegraphics[width=38mm]{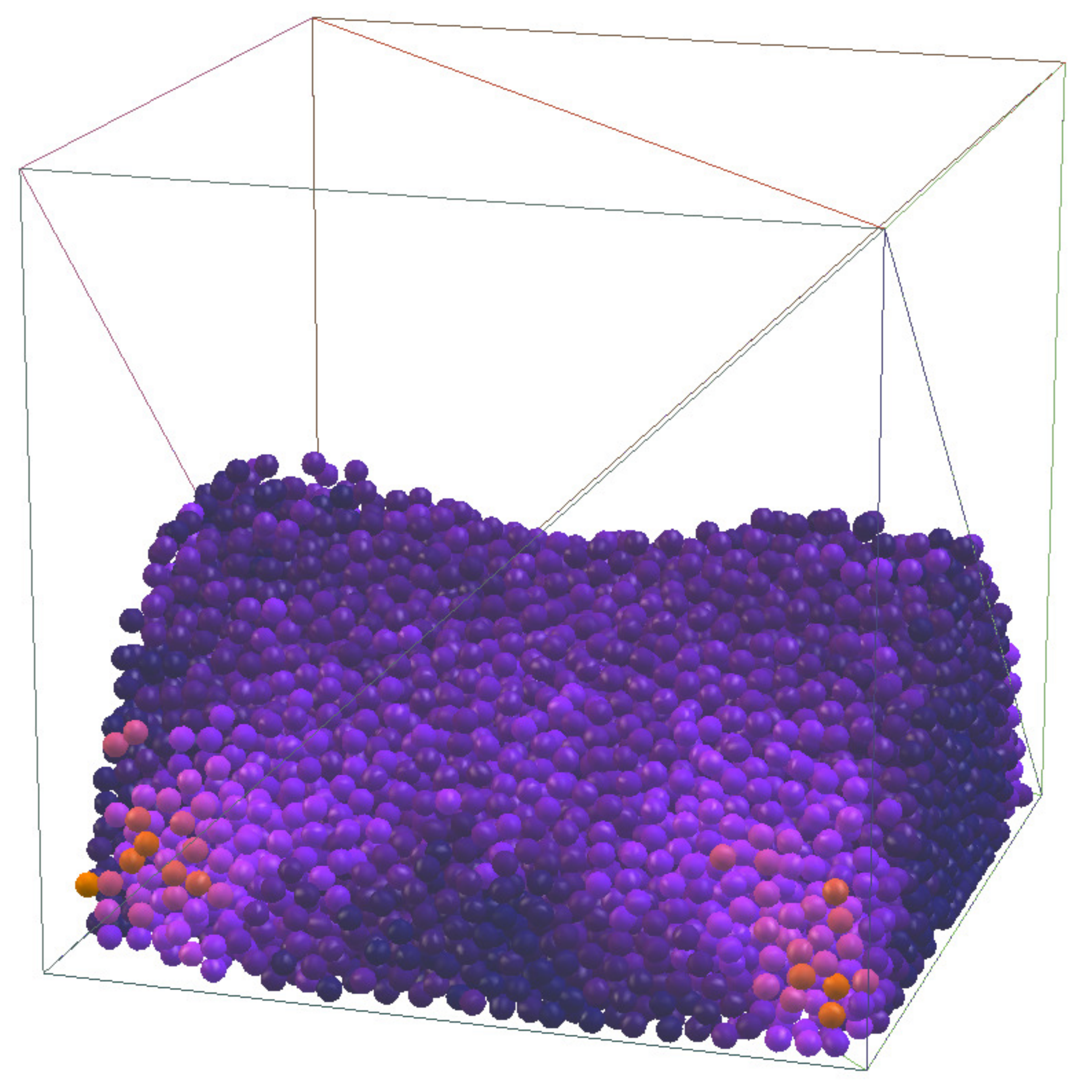}\\(c)}%
		\vtop{\includegraphics[width=38mm]{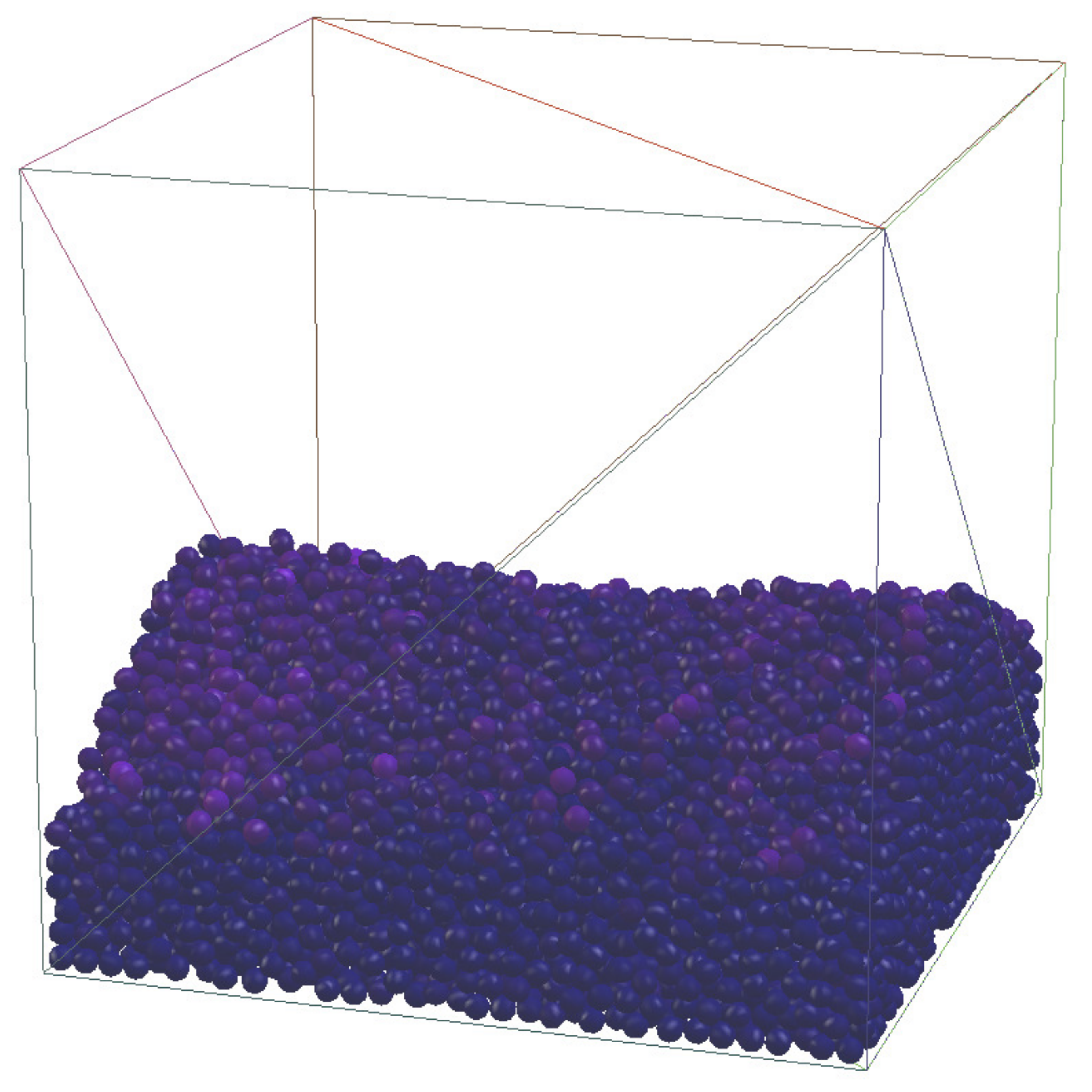}\\(d)}%
		\vtop{\setlength{\unitlength}{1.0mm}\begin{picture}(4,35)\renewcommand{\baselinestretch}{1.1}\setlength\fboxsep{0pt}%
		\put(-9,1){
		\put(0,0){\colorbox[rgb]{0.0,0.0,0.2955}{\makebox(4,3){}}}%
		\put(0,3){\colorbox[rgb]{0.3162,0.01,0.8006}{\makebox(4,3){}}}%
		\put(0,6){\colorbox[rgb]{0.4472,0.04,0.9998}{\makebox(4,3){}}}%
		\put(0,9){\colorbox[rgb]{0.5477,0.09,0.8172}{\makebox(4,3){}}}%
		\put(0,12){\colorbox[rgb]{0.6324,0.16,0.3224}{\makebox(4,3){}}}%
		\put(0,15){\colorbox[rgb]{0.7071,0.25,0}{\makebox(4,3){}}}%
		\put(0,18){\colorbox[rgb]{0.7745,0.36,0}{\makebox(4,3){}}}%
		\put(0,21){\colorbox[rgb]{0.8366,0.49,0}{\makebox(4,3){}}}%
		\put(0,24){\colorbox[rgb]{0.8944,0.64,0}{\makebox(4,3){}}}%
		\put(0,27){\colorbox[rgb]{0.9486,0.80,0}{\makebox(4,3){}}}%
		\put(0,30){\colorbox[rgb]{1.0,1.0,0.2955}{\makebox(4,3){}}}%
		}%
		\put(-15,-1.4){%
		\put(0,3){\scriptsize{0.00}}%
		\put(0,9){\scriptsize{0.04}}%
		\put(0,15){\scriptsize{0.08}}%
		\put(0,21){\scriptsize{0.12}}%
		\put(0,27){\scriptsize{0.16}}%
		\put(0,33){\scriptsize{0.20}}%
		}%
		\put(-15,36.1){\footnotesize{$E_{kin}$[J]}}
		\end{picture}\\}%
	}
\caption[Snapshots of the benchmark \texttt{yade --stdperformance -j16}]{
Snapshots of the benchmark \texttt{yade --stdperformance -j16} with \tMPFR{} 150 at different time steps:
(a)~$t=0$~s,
(b)~$t=0.3$~s (3,000 iterations),
(c)~$t=0.4$~s (4,000 iterations),
(d)~$t=0.7$~s (7,000 iterations).
The particles are colored by kinetic energy.
}
\label{fig:benchmark-view}
\end{center}
\end{figure}
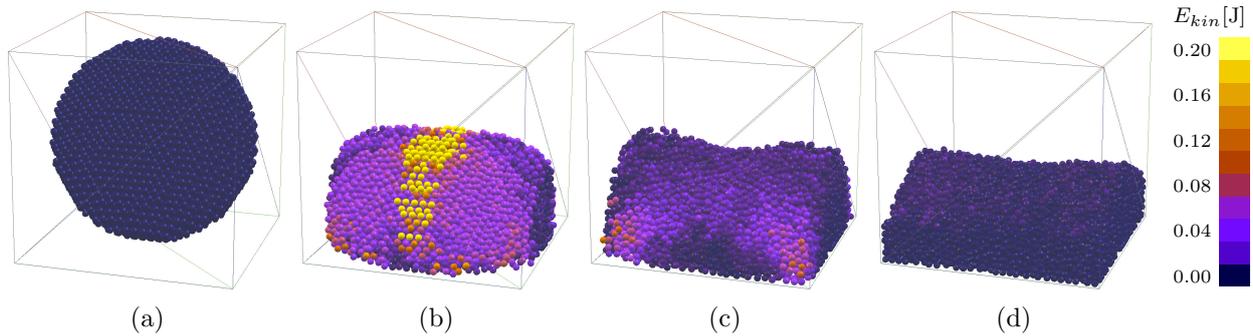

A benchmark\footnote{see: \MR{388}, \MR{491} and the
    file: \href{https://gitlab.com/yade-dev/trunk/-/blob/3e9a209234b7f23241d5f4bdef1b586056e97582/examples/test/performance/checkPerf.py\#L40}{\texttt{examples/test/performance/checkPerf.py}}}
\texttt{yade --stdperformance -j16} (16 OpenMP threads~\cite{OpenMP1,OpenMP2}) on
a PC with two \texttt{Intel E5-2687W v2 @ 3.40GHz} processors (each of the two having 8 cores resulting in a total of 16 cores or 32 threads if hyperthreading is enabled) was performed to assess performance of higher precision types.
The benchmark consists of a simple gravity deposition of spherical particles into a box, a typical simulation performed in YADE.
A spherical packing with 10,000 spheres is released under gravity within a rectangular box. The spheres are allowed to settle in the box (\Fig{benchmark-view}).
The simulation runs for 7,000 iterations and the performance is reported in terms of iterations per wallclock seconds.
This standardized test (hence \texttt{--stdperformance} in the name) is constructed in such a way that almost all the computation
happens on the C++ side, only the calculation of the wallclock time is done in Python.
Obviously doing more calculations in Python will make any script slower. Hence, any calculation in Python should be kept to a minimum.

\begin{figure}
    \centering
    $~$\\[-26mm]
    \includegraphics[height=\textheight]{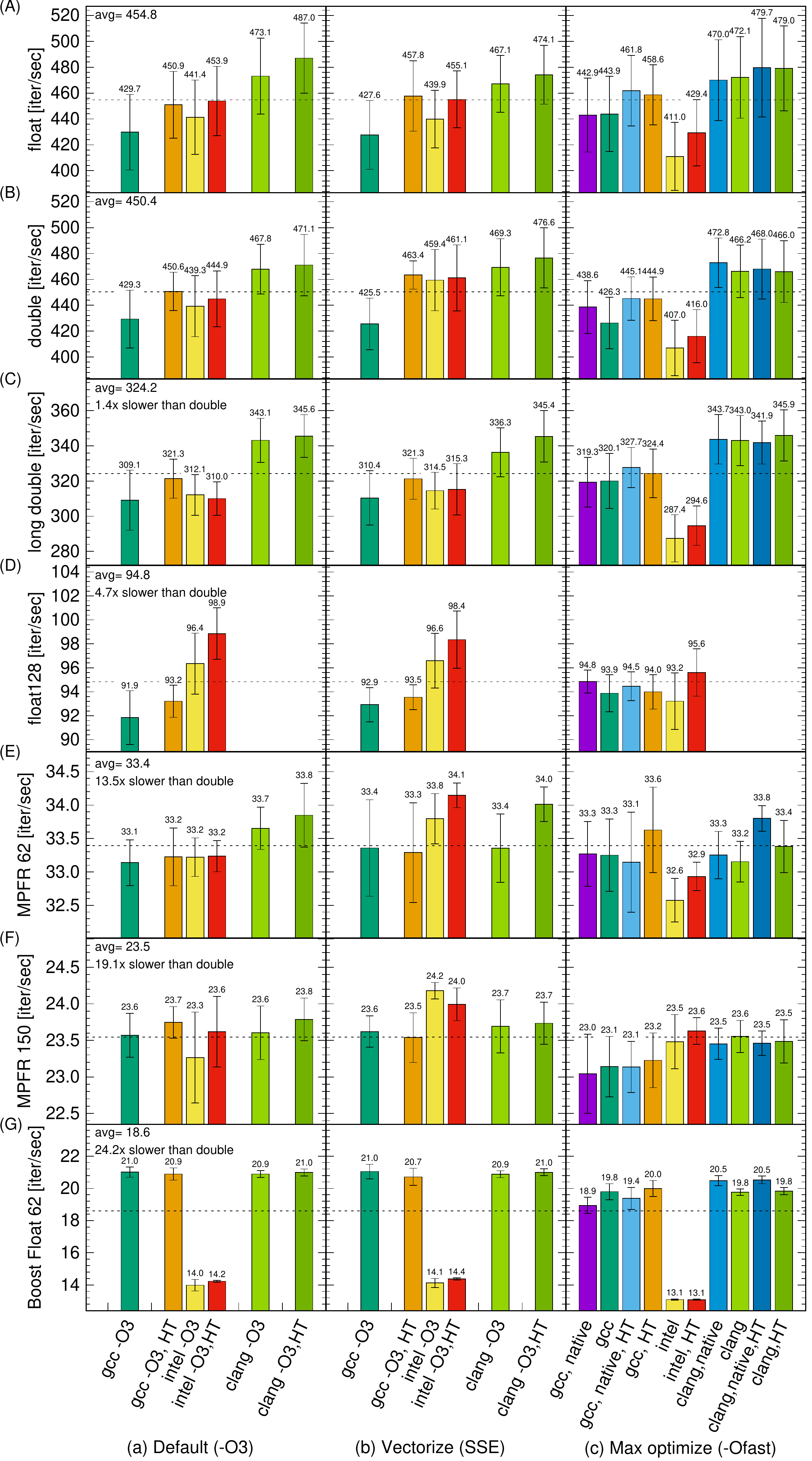}
    \caption[Benchmark results \texttt{yade --stdperformance -j16}]{Benchmark results$^\dag$ \texttt{yade --stdperformance -j16} for seven different precision types, with hyperthreading disabled$^\ddag$ and enabled (HT); for gcc version \texttt{9.3.0}, clang version \texttt{10.0.1-+rc4-1}, and intel icpc compiler version \texttt{19.0.5.281 20190815}: (a)~default \texttt{cmake} settings; (b)~with SSE vectorization enabled via \texttt{cmake -DVECTORIZE=1}; (c)~with maximum optimizations offered by the compiler (gcc, clang: \texttt{-Ofast -fno-associative-math -fno-finite-math-only -fsigned-zeros}$^\mathsection$ and additionally for native: \texttt{-march=native -mtune=native}; intel \texttt{icpc -fast}$^\mathsection$) but without vectorization.
        \\\hspace{\textwidth}$^\dag$~on a PC with two \texttt{Intel E5-2687W v2 @ 3.40GHz} processors with 16 cores and 32 threads.
        \\\hspace{\textwidth}$^\ddag$~via command \texttt{echo off > /sys/devices/system/cpu/smt/control} or by a BIOS setting (HT is also known as intel SMT).
        \\\hspace{\textwidth}$^\mathsection$~the extra three flags are used by CGAL; in intel compilers \texttt{-fast} enforces all processor model \texttt{native} optimizations.
    }
    \label{fig:benchmark}
\end{figure}

Since the benchmark results strongly depend on other processes running on the system, the test was performed at
highest process priority after first making sure that all unrelated processes are stopped (via \texttt{pkill -SIGSTOP} command).
The benchmark was \revB{repeated at least 50 times for each precision type and compiler settings.
The average calculation speed $\bar{x}$ (in iterations per seconds) was determined for each precision type and data points
not meeting the criterion $2\sigma < x_i - \bar{x} < 2\sigma$ ($\sigma$ is the standard deviation) were considered outliers.
On average 4\% of data points per bar was removed, the largest amount removed was 10\% (5 data points) on two occasions}.
\revB{Hence, each bar in~\Fig{benchmark} represents the average of at least 45 runs using 7,000 iterations each}.

A summary of all the benchmark results is shown in~\Fig{benchmark} along with the relevant standard deviations. The performance is indicated in terms of iterations per seconds.
The tests were performed\footnote{also see
    \url{https://gitlab.com/yade-dev/trunk/-/tree/benchmarkGcc}}\textsuperscript{,\ref{benchmark_intel}} for the seven different precision types
(\Fig{benchmark}A--G) listed in \Tab{benchmarked_types} for three different optimization settings: default \texttt{cmake} settings
(\Fig{benchmark}a), with SSE vectorization enabled (\Fig{benchmark}b), and with maximum optimizations offered by the compiler but without
vectorization\footnote{the test with SSE and maximum optimizations was also performed but the results were simply additive, thus they were
    not included here. Also sometimes they produced the following error due to memory alignment problems:
    \url{http://eigen.tuxfamily.org/dox-devel/group__TopicUnalignedArrayAssert.html}, because the operands of an SSE assembly SIMD instruction
    set must have their addresses to be a multiple of 32 or 64 bytes, and the compiler could not always guarantee this.} (\Fig{benchmark}c).
The lack of significant improvement in the third case (\Fig{benchmark}c) shows that the code is already well optimized and the compiler
cannot optimize it any further, except for \tldouble{} and \tfloatHTE{} types and the \texttt{gcc} compiler where a 2\% speed gain
can be observed (\Fig{benchmark}Cc and Dc versus Ca and Da).
It is interesting to note that the \texttt{clang} compiler systematically produced a code that runs about 4 to 9\% faster than the \texttt{gcc} or \texttt{icpc} compilers.
Intel compiler users should be careful, because the \texttt{-fast} switch
might result in a performance loss of around 2 to 10\%, depending on particular settings (\Fig{benchmark}c). Code vectorization (using the
SSE assembly instruction set, an experimental feature, \Fig{benchmark}b) provides about 1 to 3\% speed gain,
\revB{however this effect is often smaller than the $\sigma$ error bars}.
Enabling intel hyperthreading (HT) did not affect the results more than the standard deviation error of the benchmark.
The \tfloatHTE{} results for the intel compiler stand out with a 5\% speed gain (\Fig{benchmark}D). However, not all mathematical functions
are currently available for this precision in \texttt{icpc} and to get this test to work a crippled
branch\footnote{\label{benchmark_intel}see \url{https://gitlab.com/yade-dev/trunk/-/tree/benchmarkIntel}} was prepared for the tests with
some of the mathematical functions disabled. The missing mathematical functions\footnote{see changes in
    \href{https://gitlab.com/yade-dev/trunk/-/commit/3b07475e38173e45ab9baddb3688e7ebf93df4ef}{\texttt{MathFunctions.hpp}} in commit
    \texttt{3b07475e38} in \href{https://gitlab.com/yade-dev/trunk/-/tree/benchmarkIntel}{\texttt{benchmarkIntel}} branch} were not required
for these particular calculations to work. The \texttt{clang} compiler does not support\footnote{\label{clang_float128}see
    \url{https://github.com/boostorg/math/issues/181}} \tfloatHTE{} type yet.
The average speed difference between each precision is listed in \Tab{benchmarked_types}.
\revA{The run time increase with precision in the MPFR library is roughly $\mathcal{O}(N \log(N))$ (where $N$ is the number of digits used)
but it is application specific and strongly depends on the type of simulation performed~\cite{Isupov2020,Fousse2007}}.

\begin{table*}[ht]
    \caption{The high-precision types used in the benchmark and corresponding speed performance relative to \texttt{double}.}
    \label{tbl:benchmarked_types}
    \begin{center}
        \begin{tabular}{l r r r}
            \toprule
            Type                           & Decimal places  & Speed relative to \texttt{double}    \\
            \toprule
            \texttt{float}                 & 6       & $1.01\times$ faster   \\
            \texttt{double}                & 15      & ---                   \\
            \texttt{long double}           & 18      & $1.4\times$ slower    \\
            \texttt{boost float128}$^\dag$ & 33      & $4.7\times$ slower    \\
            \texttt{boost mpfr}$^\ddag$    & 62      & $13.5\times$ slower   \\
            \texttt{boost mpfr}            & 150     & $19.1\times$ slower   \\
            \texttt{boost cpp\_bin\_float} & 62      & $24.2\times$ slower   \\
            \bottomrule
        \end{tabular}\\
        $^\dag$~\scriptsize{except for clang which does not yet\textsuperscript{\ref{clang_float128}} support \texttt{float128}.}\\
        $^\ddag$~\scriptsize{for future comparison with \texttt{libqd-dev}, see footnote\textsuperscript{\ref{libqd_footnote}}.~~~~~~}
    \end{center}
\end{table*}

It shall be noted that currently YADE does not fully take advantage of the SSE assembly instructions (\texttt{cmake -DVECTORIZE=1})
because \texttt{Vector3r} is a three component type, while a four component class \texttt{Eigen::AlignedVector3}\footnote{four \tdouble{}
    components in \texttt{Eigen::AlignedVector3} use 256 bits which matches SSE operations, the fourth component is unused and set to zero} is
suggested in the EIGEN library but it is not completely functional yet. In the future, this class can be improved in EIGEN and then used in YADE.

\begin{table*}[ht]

\caption{Maximum error after $2\times10^7$ function evaluations expressed in terms of Units in the Last Place (ULP)$^\dag$, calculated by the absolute value of \texttt{boost::math::float\_distance}$^\ddag$ between functions from
C++ standard library or \texttt{boost::multiprecision} when compared with its respective \texttt{RealHP<4>} (having four times higher precision)$^\mathsection$.}
\label{tbl:testing_functions}
\begin{center}
\begin{tabular}{l*{7}{S}}
\addlinespace[3pt]
\toprule
& \multicolumn{7}{c}{type and number of decimal places} \\
\cmidrule(lr){2-8}
~                        & \tfloat & \tdouble & \tldouble & \tfloatHTE & \multicolumn{2}{c}{\tMPFR} & \tBBFLOAT  \\
\cmidrule(lr){6-7}
decimal places           & 6       & 15       &  18       & 33         & 62          & 150          & 62         \\
significand bits         & 24      & 53       &  64       & 113        & 207         & 500          & 207        \\
\toprule
\fun{+}                  & 0       & 0        &  0        & 0          & 0           & 0            & 0          \\
\fun{-}                  & 0       & 0        &  0        & 0          & 0           & 0            & 0          \\
\fun{*}                  & 0       & 0        &  0        & 1          & 0           & 0            & 0          \\
\fun{/}                  & 0       & 0        &  0        & 1          & 0           & 0            & 0          \\
\midrule
\fun{sin}                & 1       & 1        &  1        & 1          & 0           & 0            & 6.43e+07   \\
\fun{cos}                & 1       & 1        &  1        & 1          & 0           & 0            & 6.69e+07   \\
\fun{tan}                & 1       & 0        &  2        & 1          & 0           & 0            & 7.64e+07   \\
\fun{sinh}               & 2       & 2        &  3        & 2          & 0           & 0            & 125        \\
\fun{cosh}               & 2       & 1        &  2        & 1          & 0           & 0            & 125        \\
\fun{tanh}               & 2       & 2        &  3        & 2          & 0           & 0            & 9          \\
\midrule
\fun{asin}               & 1       & 0        &  1        & 1          & 0           & 0            & 106        \\
\fun{acos}               & 1       & 0        &  1        & 1          & 0           & 0            & 13526      \\
\fun{atan}               & 1       & 0        &  1        & 1          & 0           & 0            & 8          \\
\fun{asinh}              & 2       & 2        &  3        & 3          & 0           & 0            & 16         \\
\fun{acosh}              & 2       & 2        &  3        & 3          & 1           & 1            & 17         \\
\fun{atanh}              & 2       & 2        &  3        & 3          & 0           & 0            & 23         \\
\fun{atan2}              & 1       & 0        &  1        & 2          & 0           & 0            & 10         \\
\midrule
\fun{log}                & 1       & 1        &  1        & 1          & 0           & 0            & 17         \\
\fun{log10}              & 2       & 2        &  1        & 1          & 0           & 0            & 32         \\
\fun{log1p}              & 1       & 1        &  1        & 2          & 0           & 0            & 17         \\
\fun{log2}               & 1       & 1        &  1        & 2          & 0           & 0            & 25         \\
\fun{logb}               & 0       & 0        &  0        & 0          & 0           & 0            & 0          \\
\midrule
\fun{exp}                & 1       & 1        &  1        & 1          & 0           & 0            & 125        \\
\fun{exp2}               & 1       & 1        &  1        & 1          & 0           & 0            & 5          \\
\fun{expm1}              & 1       & 1        &  2        & 2          & 0           & 0            & 125        \\
\midrule
\fun{pow}                & 1       & 1        &  1        & 1          & 0           & 0            & 118        \\
\fun{sqrt}               & 0       & 0        &  0        & 1          & 0           & 0            & 0          \\
\fun{cbrt}               & 1       & 3        &  1        & 1          & 0           & 0            & 3          \\
\fun{hypot}              & 0       & 1        &  1        & 2          & 2           & 2            & 2          \\
\midrule
\fun{erf}                & 1       & 1        &  1        & 1          & 0           & 0            & 21         \\
\fun{erfc}               & 3       & 4        &  3        & 3          & 0           & 0            & 22496      \\
\fun{lgamma}             & 6       & 8        &  7        & 7          & 0           & 0            & 70843      \\
\fun{tgamma}             & 7       & 7        &  7        & 7          & 0           & 0            & 10661      \\
\midrule
\fun{fmod}               & 0       & 0        &  0        & 0          & 0           & 0            & 0          \\
\fun{fma}                & 0       & 0        &  0        & 1          & 0           & 0            & 1.95e+05   \\
\bottomrule

\end{tabular}\\
\begin{flushleft}
$^\dag$~\scriptsize{Also see~\cite{Kahan2006}; please note that to obtain the number of incorrect \textit{bits} one needs to take a $\log_{2}(\bullet)$ of the value in the table.}\\
$^\ddag$~\scriptsize{This test was performed with gcc version 9.3.0 and Boost library version 1.71.}\\
$^\mathsection$~\scriptsize{See file \url{https://gitlab.com/yade-dev/trunk/-/blob/master/py/high-precision/\_RealHPDiagnostics.cpp} for implementation details. This test (with fewer evaluations) can be executed using \texttt{testMath.py} and is a part of the \texttt{yade --test} suite (file \texttt{py/tests/testMath.py}, function \texttt{testRealHPErrors}).}\\
\end{flushleft}
\end{center}
\end{table*}

\section{Simulation}
\label{sec:Simulation}

\subsection{Problem description}

A simple simulation of a triple elastic pendulum system was performed to check the effect of high precision in practice.
Triple pendulums are considered highly chaotic as they provide an irregular and complex
system response~\cite{pendulum}. Numerical modeling of such systems can show the benefits of using high precision.
A single thread was used (i.e.~\texttt{yade -j1}) during the simulations to avoid numerical artifacts arising from different ordering of arithmetic operations performed by multiple threads. \revB{This allowed focusing on high precision and eliminating the non-deterministic effect of parallel calculations (e.g.~arithmetic operations performed in a different order result in a different ULP error in the last bits~\cite{Goldberg1991,Revol2014,He2001}) and to have a completely reproducible simulation}.

\begin{figure}
    \centering
    \includegraphics[width=\textwidth]{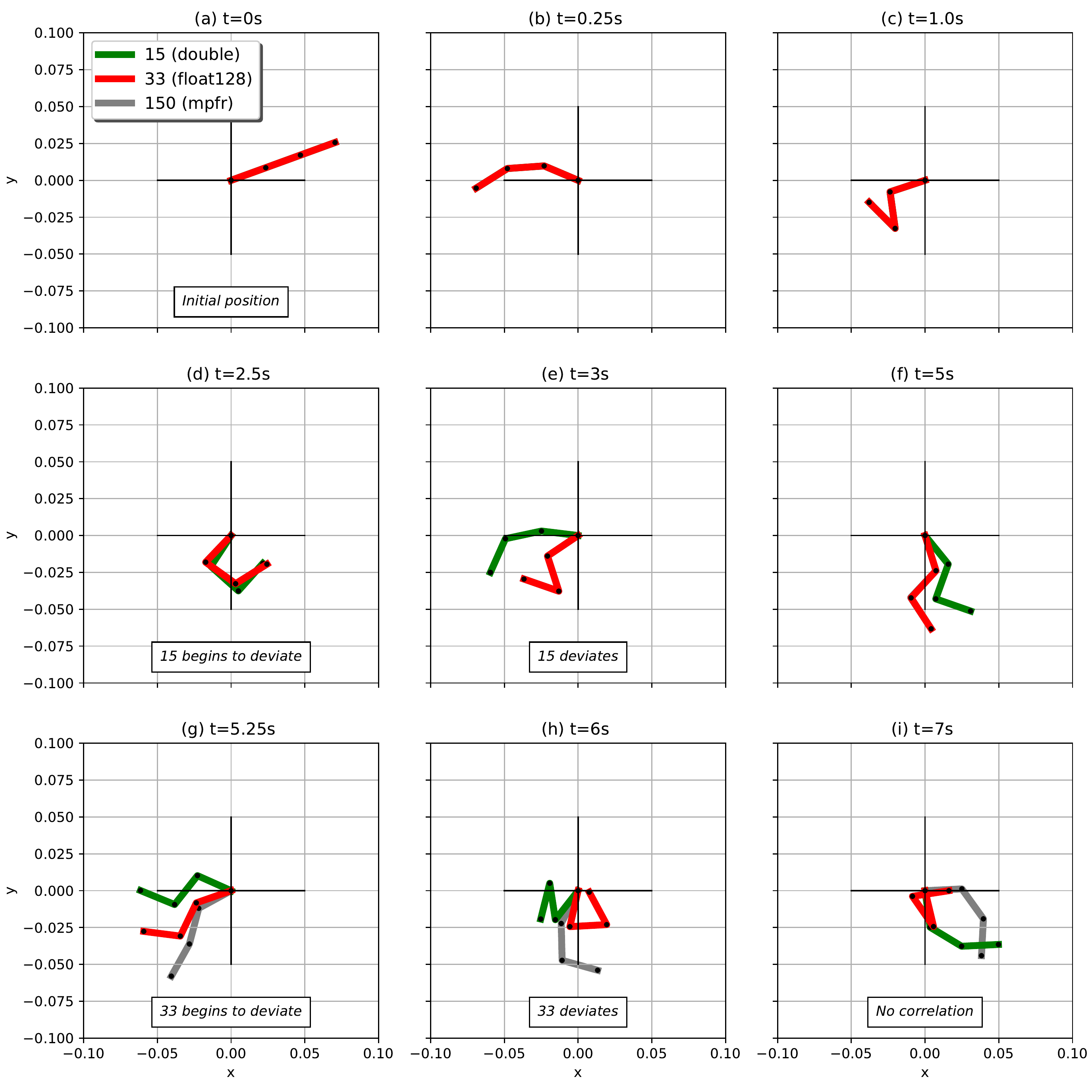}
    \caption[Triple pendulum numerical simulation]{Numerical simulation of the triple pendulum with $15$ (double), $33$ (float128) and $150$ (mpfr) decimal places.
        The snapshots are captured at the following times:
        (a)~$t=0$~s,
        (b)~$t=0.1$~s,
        (c)~$t=0.25$~s,
        (d)~$t=2.5$~s,
        (e)~$t=3$~s,
        (f)~$t=5$~s,
        (g)~$t=5.25$~s,
        (h)~$t=6$~s, and
        (i)~$t=7$~seconds.
        The lines are showing the positions of the connected rods.
        Only three precisions from \Tab{simulated_types} are shown for clarity.
    }
    \label{fig:simulation}
\end{figure}

The numerical setup of the model represents the chain consisting of three identical elastic pendulums (see attached Listing~\ref{listing1} and on \href{https://gitlab.com/yade-dev/trunk/-/tree/master/examples/triple-pendulum}{\texttt{gitlab}}).
The pendulums are represented by a massless elastic rod (a long-range normal interaction) and mass points. The latter are modeled using spheres with radius $r=0.001$~m and density $\rho=1$~kg/m$^3$, noting that the masses of the spheres are lumped into a point.
The rods are modeled by a normal interaction using cohesive interaction physics.
The length of the chain is $L=0.1$~m. Each rod is ${1/30}$~m long and the normal stiffness of the interaction is $k=100$~N/m. The strength (i.e.~cohesion) is set to an artificial high value ($10^7$~N/m$^2$) so that the chain cannot break. Hence, the behavior of the rods can be assumed purely elastic.
The initial position of the chain is $\alpha = -20\degree$ relative to the horizontal plane (see \Fig{simulation}a, $t = 0$~s).
Gravity $g=9.81$~m/s$^2$ is acting on the chain elements as they are moving.
The process was simulated with time steps $\Delta t$ equal to
$10^{-5}$~s,
$10^{-6}$~s,
$10^{-7}$~s, and
$10^{-8}$~s.
The results obtained using different precision are discussed in the following subsections. First, the evolution of the angles in the pendulum
movement are discussed with $\Delta t=10^{-5}$~s. Second, the effect of damping is shown with $\Delta t=10^{-5}$~s.
Then the effect of using various time steps $\Delta t$ is discussed. Finally, the total energy conservation is examined for various time steps.

\subsection{Pendulum movement}
\label{sec:pendulum}

Numerical damping was not used in this simulation series to avoid any energy loss and for the purity of the numerical results.
The simulations were carried out with the different precisions listed in \Tab{simulated_types} and a time step of $\Delta t=10^{-5}$~s.
Angles between the two rods were constantly monitored and saved with a period of $10^{-4}$~s for
further analysis and comparison.
After the simulation was performed and the data was gathered, the Pearson correlation coefficient \cite{stigler1989} was calculated for all data sets.
The simulation with 150 decimal places (type \tMPFR{} 150) was used as the reference solution as it has the highest number of decimal places.
The product of the two angles between the three rods was used as an input parameter for the calculation of the correlation coefficient.
The data for the correlation was placed in chunks with each having 500 elements ($10^{-4}$~s~$\times500=0.005$~s of the simulation).
Then the \texttt{scipy.stats.pearsonr} function from \texttt{SciPy} \cite{2020SciPy-NMeth} was employed for the calculation of the Pearson correlation coefficient $p$.
For further reference, the point in time when the correlation $p$ between two simulations falls below $p<0.9$ is marked as $t_s$. The latter corresponds to the time from the start of the simulation until the correlation is lost and in the following it is denoted correlation duration.

\begin{table*}[ht]
    \caption{The high-precision types used in the simulation and corresponding correlation duration $t_s$ for $\Delta t = 10^{-5}$.}
    \label{tbl:simulated_types}
    \begin{center}
        \begin{tabular}{l r r r}
            \toprule
            type                           & Decimal places  & Correlation duration $t_s$      \\
            \toprule
            \texttt{float}                 & 6   & $1.1$ seconds \\
            \texttt{double}                & 15  & $2.5$ seconds \\
            \texttt{long double}           & 18  & $3.1$ seconds \\
            \texttt{boost float128}        & 33  & $5.1$ seconds \\
            \texttt{boost mpfr}            & 62  & $9.9$ seconds \\
            \texttt{boost cpp\_bin\_float} & 62  & $9.9$ seconds \\
            \bottomrule
        \end{tabular}\\
    \end{center}
\end{table*}

\Fig{simulation} shows snapshots of the evolution of the movement for three different precisions.
At the beginning of the simulation the angles between the rods are the same. However, from a certain point in time onward the angles are starting to differ.
Indeed, at $t = 2.5$~s (\Fig{simulation}d), the green line (15 decimal places, type \tdouble) has clearly
another state compared to the other two with 33 and 150 decimal places (types \tfloatHTE{} and \tMPFR{} 150 respectively). The snapshot at $t = 5.25$~s (\Fig{simulation}g) demonstrates
the beginning of the deviation of the simulation
with 33 decimal places (type \tfloatHTE). Thereafter all the pendulums are moving very differently and no correlation is observed.
It can be concluded that using higher precision increases the time when accurate calculation results are obtained which is also reflected by the correlation duration $t_s$ listed in \Tab{simulated_types}.

\begin{figure}
    \centering
    \includegraphics[width=1.0\textwidth]{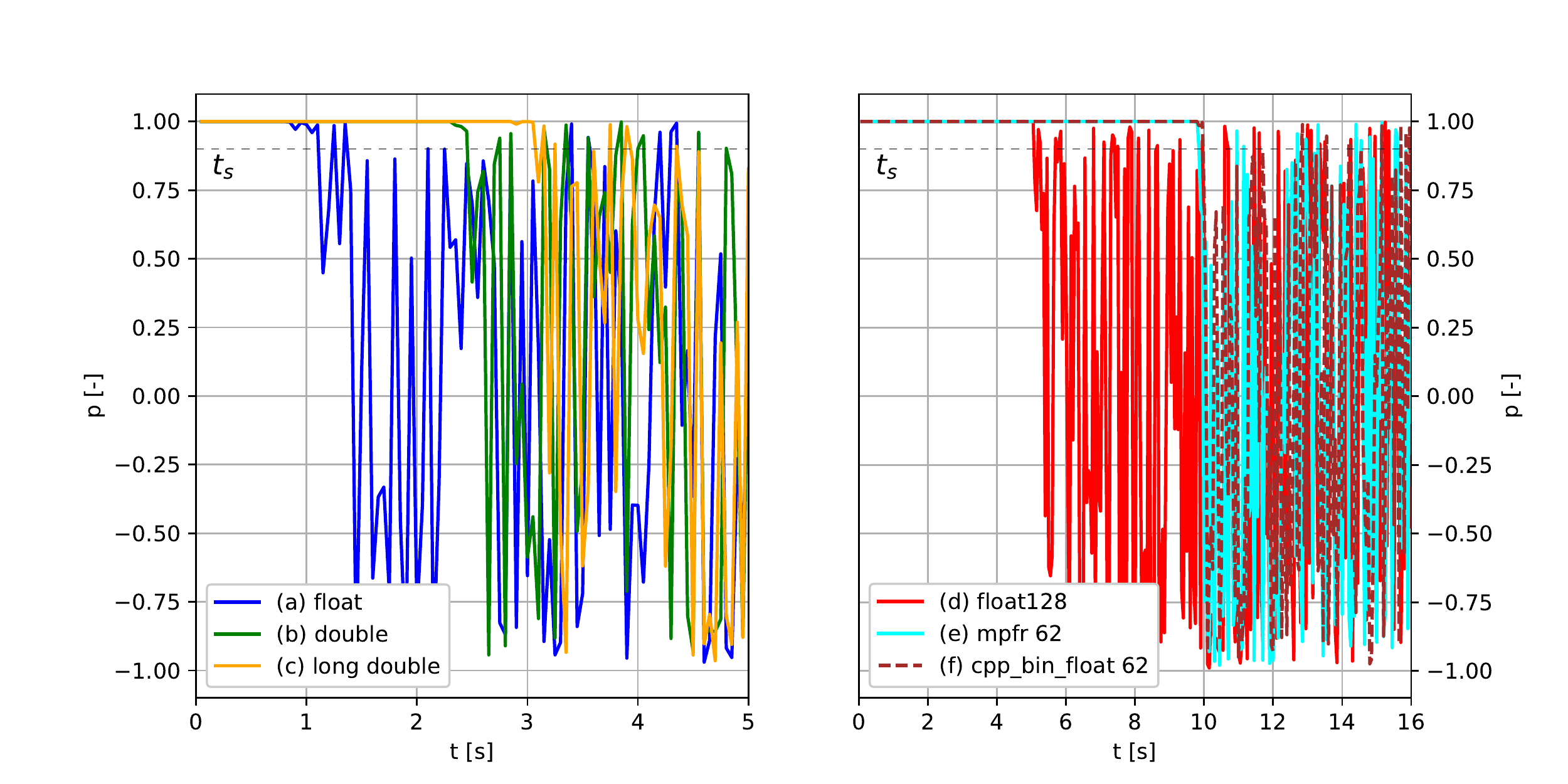}
    \caption[Pearson correlation coefficient $p$ as a function of time $t$]{Pearson correlation coefficient $p$ as a function of time $t$
        between the results obtained with \tMPFR{} 150 and the following different precisions:
        (a)~\tfloat{};
        (b)~\tdouble{};
        (c)~\tldouble{};
        (d)~\tfloatHTE{};
        (e)~\tMPFR{} 62;
        (f)~\tBBFLOAT{} 62.
        Note that the timescales on both figures are different.
        The black dashed lines mark the threshold $p=0.9$ which is used to calculate the correlation duration $t_s$.
    }
    \label{fig:correlation-simulations}
\end{figure}

\Fig{correlation-simulations} presents the correlation coefficient as a function of time. The graphs provide a more
accurate representation of the point in time when the correlation disappears.
One can see that there is a positive linear correlation with $p=1$ at the beginning of the simulations. This means initially the rods are moving identically. The lowest precision curve \tfloat{} is starting to jump between correlation values of $p=1$ and $p=-1$ at around $t=1.1$~s (\Fig{correlation-simulations}a), which means that no visible correlation is observed any more. For all the higher precision simulations the drop off happens later and progressively with increasing precision as summarized in \Tab{simulated_types}.

Type \tdouble{} and \tldouble{} have a correlation of $p<0.9$ after $2.5$~s (\Fig{correlation-simulations}b, comparing 15 with 150 decimal places) and $3.1$~s (\Fig{correlation-simulations}c, 18 vs.~150 decimal places) respectively.
The same tendency is seen from Boost \tfloatHTE{} type (\Fig{correlation-simulations}d, 33 vs.~150 decimal places) which deviates at
approximately $5.1$~s. Both simulations with 62 decimal places start to deviate at around $9.9$~s.
This clearly demonstrates that the level of precision, i.e.~the number of decimal places,
has an influence on the accuracy of the simulation results.
Sometimes the decrease of the correlation happens suddenly and sometimes it starts to decrease slowly before it decreases rapidly.
It can clearly be seen that the simulations with higher precision are showing better results and are closer to the reference solution calculated with 150 decimal places. Nevertheless, higher precision requires much more time for the simulation and more computing resources.

\subsection{Effect of damping}

To study the importance of precision in combination with other parameters, the same simulations as in Section~\ref{sec:pendulum} were carried out with a
numerical damping coefficient equal to $5\cdot10^{-3}$.
Numerical damping is generally applied to dissipate energy. In this particular test case, numerical
damping can be interpreted as the slowing down of the pendulum oscillation. The global damping mechanism was used, as
described in the original DEM publication of Cundall and Strack~\cite{CundallStrack1979}. Global damping acts on the absolute
velocities of the simulation bodies and is implemented in the
\href{http://yade-dem.org/doc/yade.wrapper.html?highlight=newtonintegrator\#yade.wrapper.NewtonIntegrator}{\texttt{NewtonIntegrator}} class in
the source code of YADE. Global damping slows down all affected bodies based on their current velocities.

The damping coefficient was chosen so that an effect of different precisions can be seen on the whole system. If the damping coefficient is too
high ($> 10^{-1}$), the system loses its whole energy very quickly and no visible differences are seen. Too small
damping coefficients ($< 10^{-3}$) lead to very slow energy dissipation.

\begin{figure}
    \centering
    \includegraphics[width=0.5\textwidth]{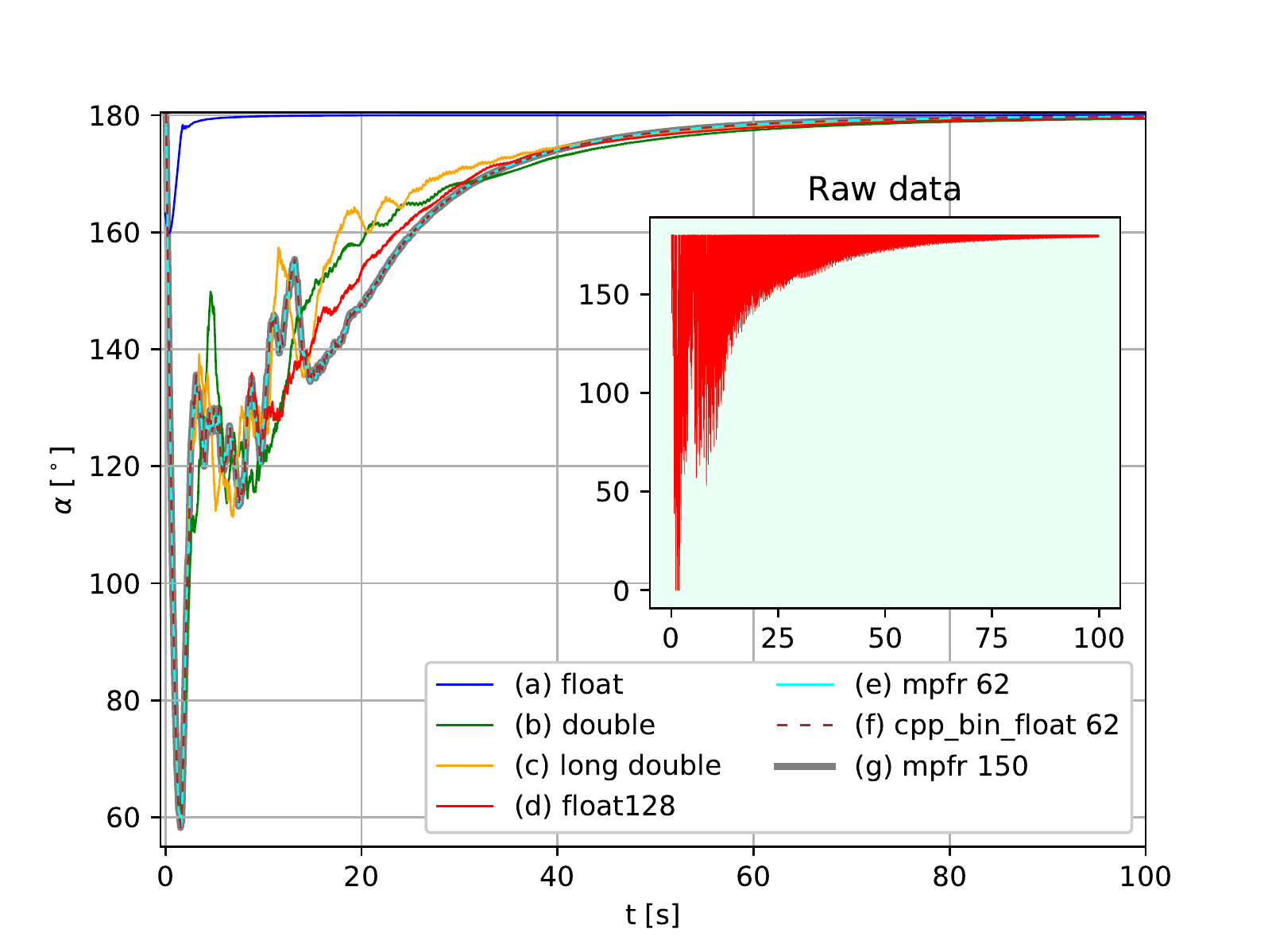}
    \caption[Angle development between first and second rod]{Development of angle $\alpha$ between first and second rods as a function of time $t$ for different precisions with a global damping coefficient equal to $5\cdot10^{-3}$.\
        Curves are showing smoothed data for better visibility (main plot) and raw data for the inset;
        (a)~\tfloat{};
        (b)~\tdouble{};
        (c)~\tldouble{};
        (d)~\tfloatHTE{};
        (e)~\tMPFR{} 62;
        (f)~\tBBFLOAT{} 62.
        (g)~\tMPFR{} 150.
    }
    \label{fig:angle-damping}
\end{figure}

\Fig{angle-damping} shows the development of the angle between the first and the second rod of the pendulum during the simulation
with a global damping coefficient equal to $5\cdot10^{-3}$.
One can clearly see the differences in simulation results based on different precisions. The \tfloat{} precision simulation
(blue curve, \Fig{angle-damping}a) indicates the largest deviation from all other simulations. Higher precisions gradually provide
results that are closer to the simulation with the highest precision (boost \tMPFR{} 150 decimal places).

Since the angle between rods is oscillating rapidly, as can be seen
on the inset in \Fig{angle-damping},
the raw data was smoothed using the Savitzky–Golay filter~\cite{savitzky64} for better visibility.
The filter removes most of the noise, and there are no visible differences between \tBBFLOAT{}, \tMPFR{} 62 and \tMPFR{} 150. The three curves are basically overlapping each other.

\subsection{Effect of time step $\Delta t$}
\label{sec:timestep}

\begin{figure}
    \centering
    \includegraphics[width=0.5\textwidth]{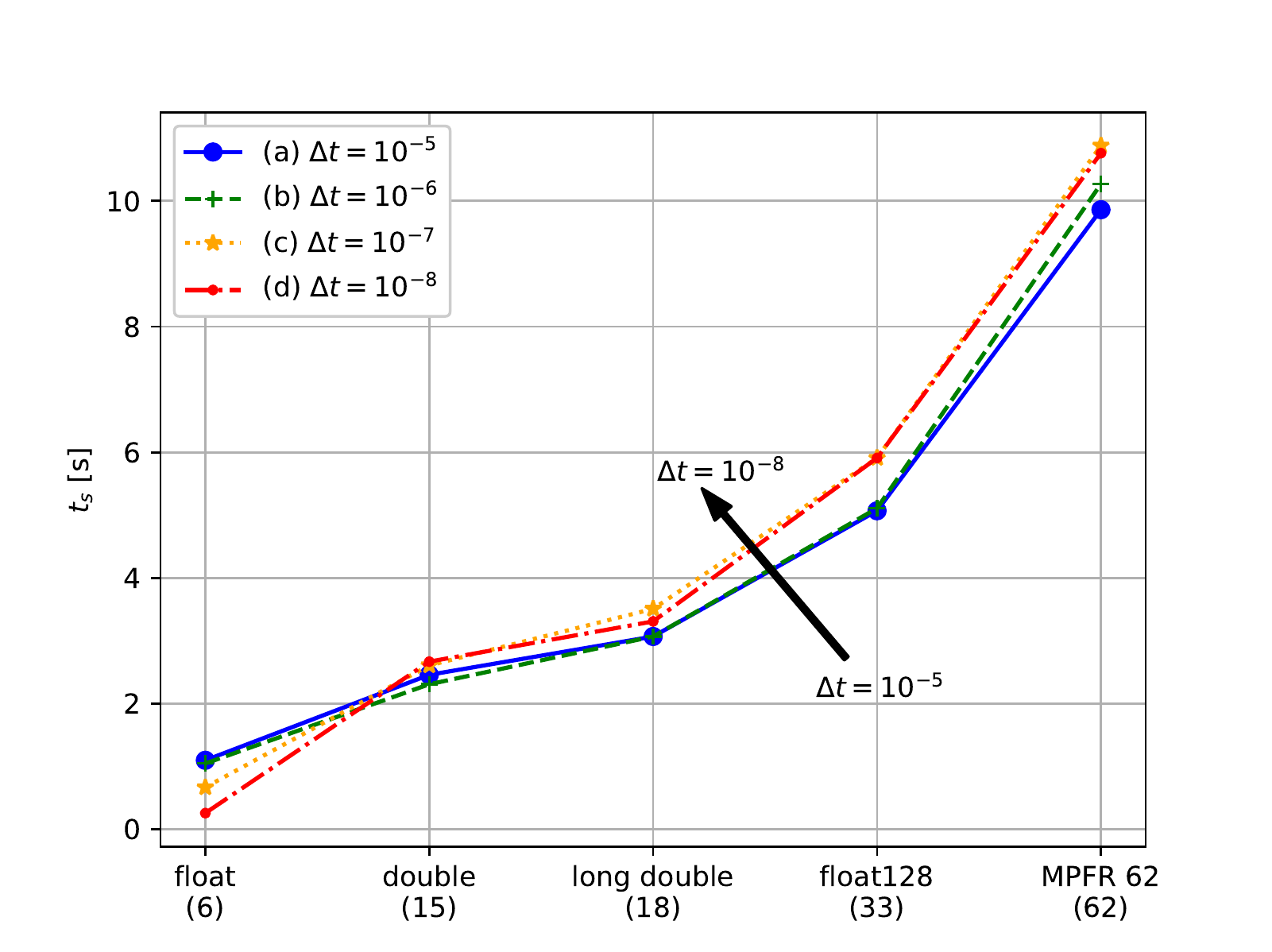}
    \caption[Correlation duration $t_s$]{Correlation duration $t_s$ as a function of different precisions and a time step $\Delta t$.
    Each curve is compared to the \tMPFR{} 150 simulation using the same $\Delta t$;
    (a)~$\Delta t=10^{-5}$~s;
    (b)~$\Delta t=10^{-6}$~s;
    (c)~$\Delta t=10^{-7}$~s;
    (d)~$\Delta t=10^{-8}$~s.
    }
    \label{fig:perfect-correlation-duration-dt}
\end{figure}

The time step $\Delta t$ is one of the most important parameters influencing the simulation results. Hence, simulations with time step values of
$10^{-5}$~s,
$10^{-6}$~s,
$10^{-7}$~s, and
$10^{-8}$~s
were performed with different precisions. The values for the correlation duration were recorded
and plotted in \Fig{perfect-correlation-duration-dt} for all precisions and time steps considered.
It can be clearly seen that the correlation duration increases with increasing precision. This highlights once more that higher precision is required for higher confidence.
It can also be seen that generally the correlation duration increases with decreasing time step. This is due the fact that a smaller time step results in a smaller integration error (e.g. the leapfrog integration scheme used in YADE
has the error proportional to $\Delta t^2$ per iteration).
This tendency is also marked on the figure with a black arrow pointing in the direction of the smaller time step $\Delta t$.
The opposite result is observed for \tfloat{}. This is because 6 decimal places are not enough to work with $\Delta t=10^{-8}$~s.
It shall be noted that the implementation of a more precise time integration scheme is out of the scope of the current paper.
In the current symplectic leapfrog integration scheme, as implemented in the present version of
\href{http://yade-dem.org/doc/yade.wrapper.html?highlight=newtonintegrator\#yade.wrapper.NewtonIntegrator}{\texttt{NewtonIntegrator}}, the positions and velocities
are leapfrogging over each other.
There is no jump-start option which would allow to start the simulation with both position and velocity defined at $t=0$.
This means that the initial velocities are declared at $t=-\frac{\Delta t}{2}$ and initial positions at $t=0$ or
initial velocities are declared at $t=0$ and initial positions at $t=\frac{\Delta t}{2}$. Which of these two it is, is only a
formal choice\footnote{see also: \url{https://gitlab.com/yade-dev/trunk/-/merge_requests/555\#note_462560944} and \href{https://gitlab.com/yade-dev/trunk/-/blob/e43c3fb4b90485a879051b0a60dd2dd1c1b2a357/scripts/checks-and-tests/checks/checkGravity.py\#L51}{\texttt{checks/checkGravity.py}}},
since both interpretations are valid.
Therefore a more detailed comparison between the time steps is not carried out as the initial conditions for each simulation with a different $\Delta t$ differ slightly,
i.e.~the starting velocity
declared in the script is interpreted as being defined at $t=-\frac{10^{-5}}{2}$ or at $t=-\frac{10^{-8}}{2}$, thus resulting in slightly different simulations\textsuperscript{\ref{future_integrators}}.

\subsection{Energy conservation}
\label{sec:energy}

In the following, the total energy in the system is analyzed for different precisions and time step values of
$10^{-5}$~s,
$10^{-6}$~s,
$10^{-7}$~s, and
$10^{-8}$~s.
No numerical damping is considered.
The total energy in the system is calculated as the sum of the elastic energy in the interactions and the kinetic and potential energy of the mass points (i.e.~spheres).
As already pointed out in the previous section, the symplectic leapfrog integration scheme is used.
This means that velocities and positions are not known at the same time. Hence, the velocities needed for an accurate calculation of the kinetic energy are taken as an
average of the velocities from the current and the next iteration.
All numerical results are compared to the reference solution which was calculated using 150 decimal places, similar as in the previous sections.

\begin{figure}
    \centering
    \begin{subfigure}[b]{0.49\textwidth}
        \centering
        \includegraphics[width=\textwidth]{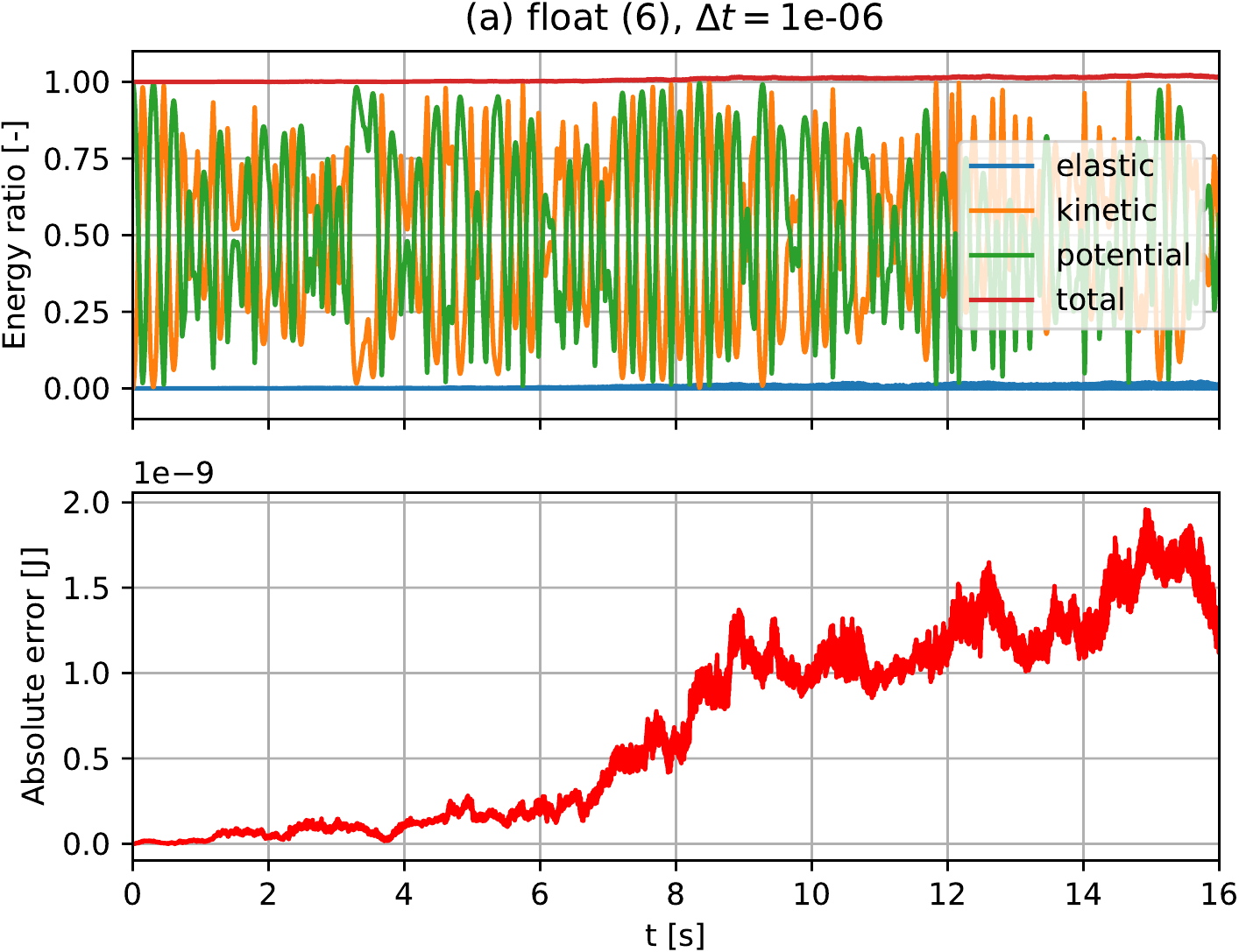}
    \end{subfigure}
    \hfill
    \begin{subfigure}[b]{0.49\textwidth}
        \centering
        \includegraphics[width=\textwidth]{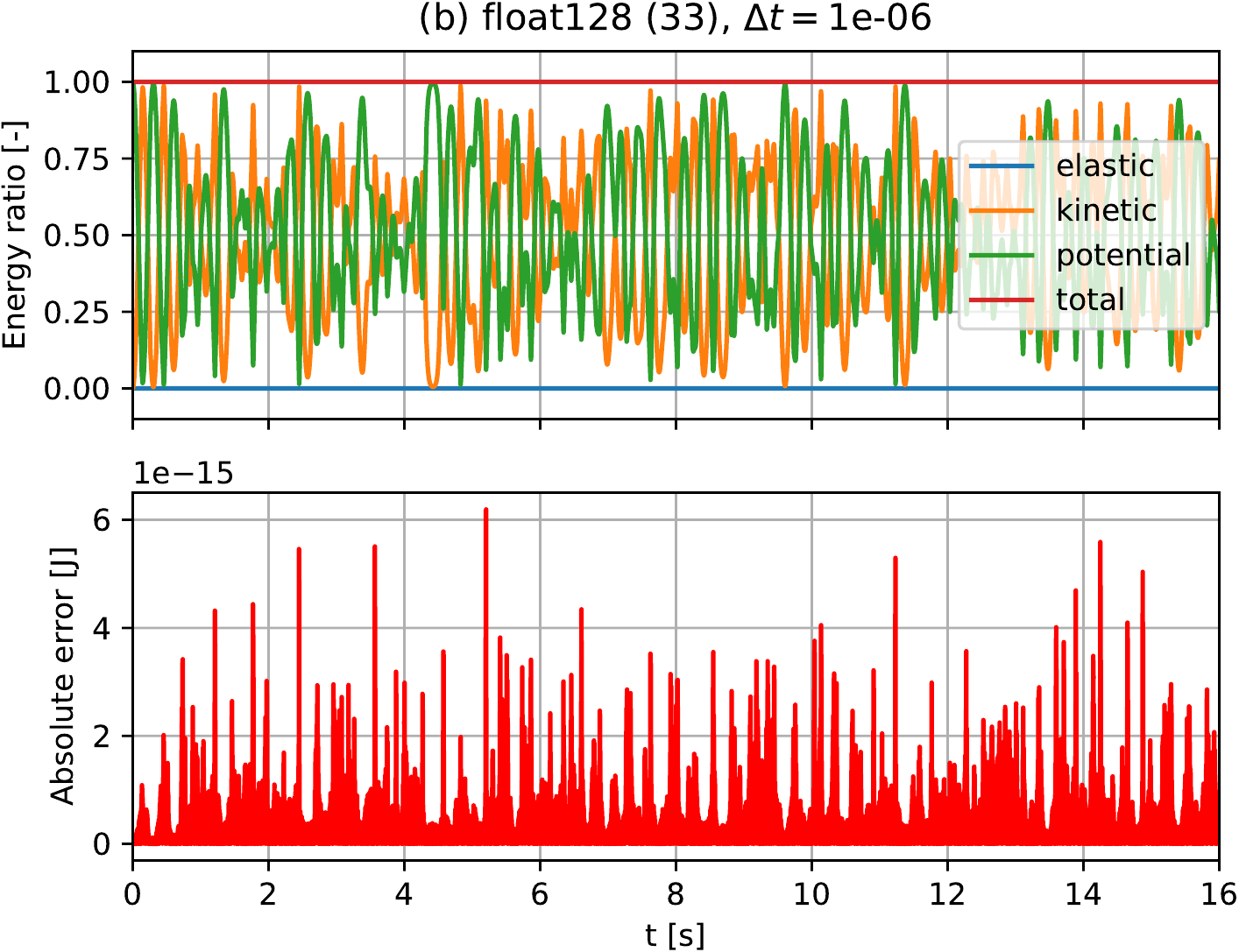}
    \end{subfigure}
    \caption[]{Evolution of energy balance plotted as energy ratios and corresponding absolute error compared to the \tMPFR{} 150 simulation for precisions (a)~\tfloat{} and (b)~\tfloatHTE{}.}
    \label{fig:energy-balance}
\end{figure}

\Fig{energy-balance} shows two typical results obtained from the study using a time step of $\Delta t=10^{-6}$~s. The results obtained with the other time steps have a similar trend and are not included for brevity.
The top graphs show the evolution of the energy balance where each energy component is divided by the total reference energy to give an energy ratio.
The total reference energy is calculated using 150 decimal places.
The total energy ratio should be equal to 1 throughout the simulation.
The bottom graphs show the absolute error in total energy calculated as the absolute difference between the total energy given by a specific precision and
the constant total energy calculated using 150 decimal places.

The results obtained using \tfloat{} are depicted in \Fig{energy-balance}a.
It can be seen that the absolute error in total energy starts to increase drastically after about 4~s.
This is much later than the correlation duration.
From the energy plot it can also be seen that energy is continuously added to the system from this time onward. This makes the simulation not only incorrect but also unstable.
A different observation can be drawn from \Fig{energy-balance}b where the results of a typical simulation with \tfloatHTE{} are shown.
It can be seen that the energy balance is stable and the absolute error is several order of magnitudes smaller. In addition, the absolute error does not have an increasing trend. Instead, it bounces around, i.e.~it increases initially and decreases thereafter, and never goes above a certain threshold value.
This is also a reflection of the symplectic leapfrog integration scheme.
It should be noted that the results for \tdouble{}, \tldouble{}, \tMPFR{}~62 and \tBBFLOAT{}~62 are very similar to \Fig{energy-balance}b and, hence, not shown for brevity.

\begin{figure}
    \centering
    \includegraphics[width=0.5\textwidth]{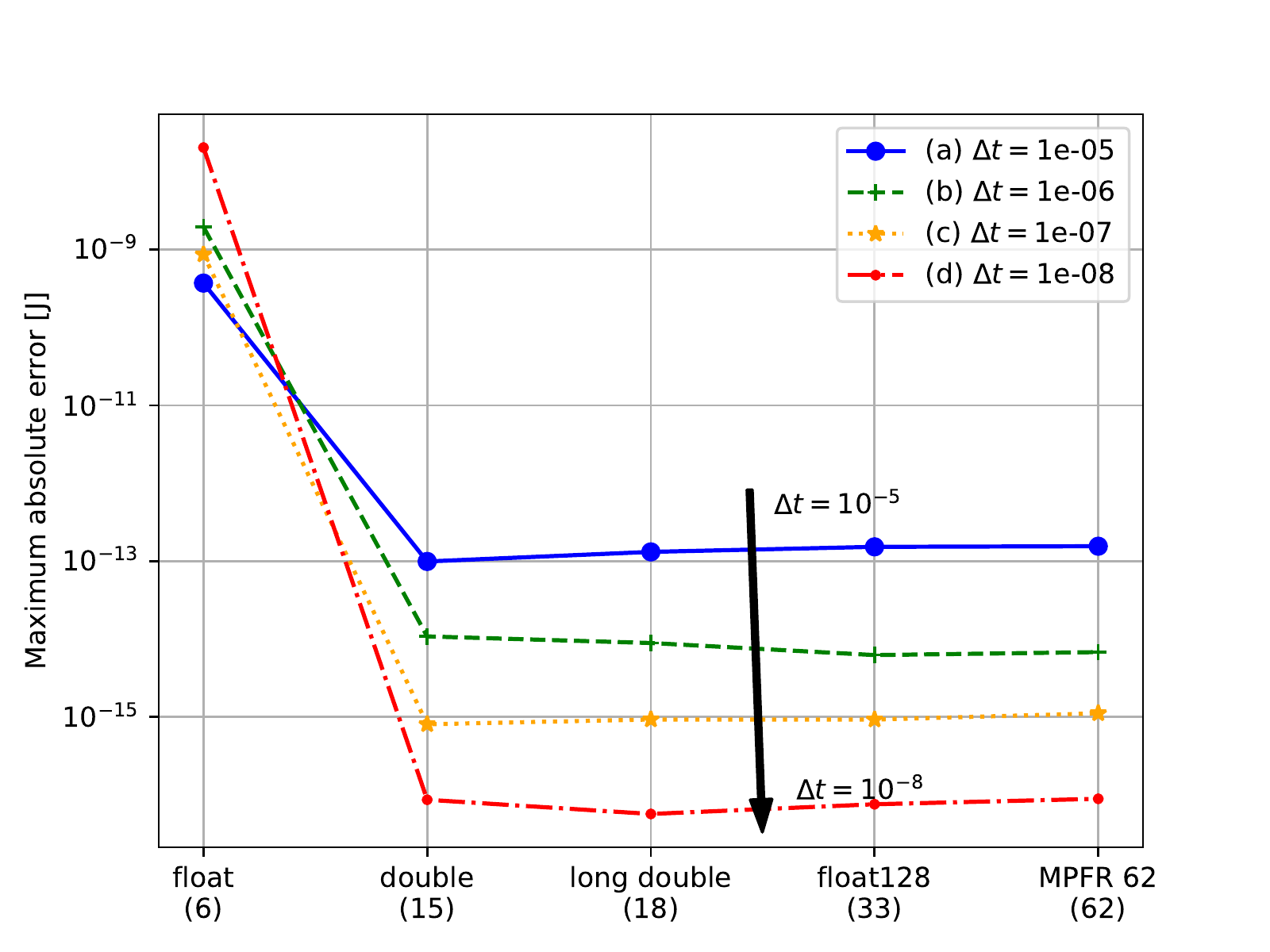}
    \caption[Absolute error in energy balance]{Maximum absolute error in energy balance during the first 16~s as a function of different precisions and for different time steps $\Delta t$.
    Each curve is compared to the \tMPFR{} 150 simulation using the same $\Delta t$;
    (a)~$\Delta t=10^{-5}$~s;
    (b)~$\Delta t=10^{-6}$~s;
    (c)~$\Delta t=10^{-7}$~s;
    (d)~$\Delta t=10^{-8}$~s.
    Note that the axis for the maximum absolute error is in log scale.
    }
    \label{fig:error-dt}
\end{figure}

\Fig{error-dt} summarizes the results for all precisions and all time steps. As noted previously, a detailed comparison between different time steps does not make sense because of the different initial velocities.
Nevertheless, a qualitative comparison is valid. It can be seen that the maximum absolute error in energy balance for \tfloat{} is many orders of magnitudes larger than for the other precisions (note that the vertical axis uses logarithmic scale).
The data also indicates that the error is almost constant for all other precisions.
This clearly highlights the effectiveness and reliability of the symplectic leapfrog integration scheme implemented in YADE.
However the error is many orders of magnitude larger than the ULP error of the higher precision types.
For example to achieve maximum absolute error of $10^{-30}$ [J] for \tfloatHTE{}, further decreasing the time step is not practical.
Different approaches, such as higher order symplectic methods, have to be employed~\cite{Omelyan2002,Omelyan2006}\textsuperscript{\ref{future_integrators}}.
Like in previous section, a smaller time step results in a smaller absolute error (this tendency is indicated by the black arrow), except for \tfloat{} where 6 decimal places are not enough to work with the smaller time steps.

\section{Conclusions and future perspectives}

The obtained results show that using high precision has a pronounced influence on the simulation results and the
calculation speed. Higher precisions provide more accurate results and reduce
numerical errors which in some fields can be beneficial.
It can also be concluded that high precision is essential for research of highly chaotic systems (\Fig{correlation-simulations}).
Nevertheless, increasing the number of decimal places in the code leads to a higher CPU load and raises
the calculation times (\Fig{benchmark}, \Tab{benchmarked_types}).

Updating an existing software with a large codebase to bring the flexibility of arbitrary precision can be a challenging and error-prone process which might require drastic refactoring.
A good test coverage of the code (unit and integration tests: the Continuous Integration pipeline in YADE) is highly recommended before beginning of
such refactoring to ensure code integrity~\cite{Bellotti2021}. Also the employment of AddressSanitizer~\cite{asan} is highly desirable to prevent heavy memory errors such as heap corruptions, memory leaks, and out-of-bounds accesses.

Simulations with the triple pendulum show that the results are starting to be different after a few seconds of the
simulation time because of different precisions.
The higher the precision the longer the results remain true to the highest precision tested.
The same effect, within each precision (\Fig{perfect-correlation-duration-dt} and \Fig{error-dt}), occurs when using smaller time steps $\Delta t$,
because it has a smaller time integration error per iteration.
Applying damping can significantly smooth this effect (\Fig{angle-damping}).

The new high-precision functionality added to YADE does not negatively affect the \revB{existing} computational performance \revB{(i.e.~simulations with \tdouble{} precision)}, because the choice of precision is done at compilation time
and is dispatched during compilation via the C++ static polymorphism template mechanisms~\cite{Stroustrup2014,Vandevoorde2017}.

\bigskip

\noindent
Concluding this work, the main modules of YADE now fully support two aspects of arbitrary precision (see \Tab{supported_packages} and \Fig{dependencies}):

\begin{enumerate}
    \item Selecting the base precision of \tReal{} from \Tab{supported_types} (which is an alias for \texttt{RealHP<1>}).
    \item Using \texttt{RealHP<N>} in the critical C++ sections of the numerical algorithms (Section~\ref{sec:RealHP})\textsuperscript{\ref{precision_benefits}}.
\end{enumerate}

\noindent
These new arbitrary precision capabilities can be used in several different ways in the management of numerical error~\cite{Bailey2005c}:

\begin{enumerate}
    \setcounter{enumi}{2}

    \item To periodically test YADE computation algorithms to check if some of them are becoming numerically sensitive.
    \item To determine how many digits in the obtained intermediate and final results are reliable.
    \item To debug the code in order to find the lines of code which produce numerical errors, using the method described in details in chapter 14 of~\cite{Kahan2006}.
    \item To fix numerical errors that were found by changing the critical part of the computation to use a higher precision type like \texttt{RealHP<2>} or \texttt{RealHP<4>} as suggested in~\cite{Kahan2004}.
\end{enumerate}

\noindent
The current research focus is to:

\begin{enumerate}
    \setcounter{enumi}{6}

    \item Add quantum dynamics calculations to YADE using the time integration algorithm which can have the error smaller than the numerical
          ULP error of any of the high-precision types: the Kosloff method~\cite{Kosloff1984,Kosloff1997,KosloffSchaefer2017,TalEzer2012} based on the rapidly
          converging Chebychev polynomial expansion of an exponential
          propagator\footnote{Since that algorithm is Taylor--free, there is no meaning to the term ``order of the method''~\cite{TalEzer2012,KosloffSchaefer2017}.}.
    \item Add unit systems support, because there are also software errors related
          to unit systems, for example in 10 November 1999 the NASA's Mars Climate Orbiter was lost in space because of mixing SI and imperial
          units~\cite{Stephenson1999,Peterson1996,Huckle2019}.
\end{enumerate}

\noindent
Possible future research avenues, opened by the present work, include:

\begin{enumerate}
    \setcounter{enumi}{8}

    \item Add more precise time integration algorithms. The problems mentioned in Section~\ref{sec:timestep} are well known.
          Albeit symplectic integrators are particularly good~\cite{Quispel1996,Sussman1992},
          a better time integration method with smaller $\Delta t$ will be able to fully take advantage of the new high-precision capabilities (Section~\ref{sec:energy} and \Fig{error-dt}).
          There are three possible research directions:

          \begin{enumerate}

              \item Use the Boost Odeint library~\cite{Boost2020} with higher order methods\textsuperscript{\ref{precision_benefits},}\footnote{\label{future_integrators}see also:
                        \href{https://www.boost.org/doc/libs/1_75_0/libs/numeric/odeint/doc/html/boost_numeric_odeint/getting_started/overview.html}{Boost Odeint}, \ISSUE{171},
                        \MR{555} and \MR{557} where \href{https://gitlab.com/yade-dev/trunk/-/blob/master/scripts/checks-and-tests/checks/checkGravityRungeKuttaCashKarp54.py}{\texttt{RungeKuttaCashKarp54Integrator}} allowed to set error tolerance of 147 correct decimal places if \tMPFR{} 150 is used.}.

              \item Use the work on time integrators \revA{by Omelyan et al.}~\cite{Omelyan2002,Omelyan2006}.
			\revA{These approaches suggests that it is potentially possible to decrease the run time more than 50-fold with the same computation effort
			upon switching to \tldouble{}, \tfloatHTE{} or higher types.
			The algorithms focus on reducing the truncation errors and eliminating errors introduced by the computation of forces.
			Such a smaller error allows to use a larger time step which will more than compensate the speed loss due to high-precision calculations.
			Of course, the standard considerations for the time step~\cite{Burns2016,Omelyan2011} would have to be re-derived.}
              \item Investigate whether the exponential propagator approach presented in~\cite{KosloffSchaefer2017} or in~\cite{Orimo2018,Scrinzi2021}
                    could be used in YADE as a general solution for ODEs, similarly to~\cite{Omelyan2006,Hochbruck1999,Caliari2009,Quispel1996,McLachlan1998,Fahs2012}, regardless if that is a classical
                    or a quantum dynamics system.
          \end{enumerate}

    \item Enhance all auxiliary modules of YADE for the use of high precision\footnote{\label{supported_modules}for the full up-to-date list of supported modules see: \url{http://yade-dem.org/doc/HighPrecisionReal.html\#supported-modules}}.

    \item \revB{Use the interval computation approach to reduce problems with numerical reproducibility in parallel computations by using \texttt{boost::multiprecision::mpfi\_float} as the backend for \texttt{RealHP<N>} type~\cite{Revol2014,Merlet2007}}. 

    \item Use different rounding modes to run the same computation for more detailed testing of numerical algorithms~\cite{Kahan2006}.

\end{enumerate}

Overall, based on the presented work, the architecture of YADE now offers an opportunity to adjust its precision according to the needs of its
user. A wide operating system support and simple installation procedure enable forming multidisciplinary teams for computational physics
simulations in the Unified Science Environment (USE)~\cite{McCurdy2002}.
This will expand the spectrum of tasks that can be solved, improve the results and reduce numerical errors.
Of course, this option not only complicates the architecture and the source code, but also imposes a restriction on the choice of
a programming language.

\section*{Acknowledgements}

The authors would like to acknowledge the partial support of the COST action CA18222 ``Attosecond Chemistry'' and the Australian Research Council (DP190102407).
The support of the Institute for Mineral Processing Machines and Recycling Systems Technology at TU Bergakademie Freiberg for providing access to computing facilities is also acknowledged.
In addition, the authors thank the Department of Building Structures and Material Engineering, Faculty of Civil and Environmental Engineering, at Gdańsk University of Technology for hosting the gitlab Continuous Integration (CI) pipeline for YADE.
\revB{The authors would also like to thank the two anonymous reviewers whose suggestions helped improve this manuscript.}

\bibliographystyle{elsarticle-harv}
\bibliography{yadeHP}

\clearpage
\begin{lstlisting}[numbers=left,label=listing1,caption={The triple pendulum simulation script},language=Python]
# The script is tested with yade_2021.01a
from yade import plot, qt, math

# initialize all floating point variables with Real(arg) to avoid precision loss
from yade.math import toHP1 as Real
# after release yade_2021.01a math.toHP1 has an alias Real, same with radiansHP1
if(math == sys.modules['math']): raise RuntimeError("Python math obscures yade.math")

### set parameters ###
L = Real('0.1')      # length [m]
n = 4                # number of nodes for the length [-]
r = L/100            # radius [m]
g = Real('9.81')     # gravity
inclination = math.radiansHP1(20)  # Initial inclination of rods [degrees]
color = [1, 0.5, 0]  # Define a color for bodies

O.dt = Real('1e-05') # time step
damp = Real('1e-1')  # damping. It is interesting to examine  damp = 0

O.engines = [  # define engines, main functions for simulation
    ForceResetter(),
    InsertionSortCollider([Bo1_Sphere_Aabb()],
                          label='ISCollider', avoidSelfInteractionMask=True),
    InteractionLoop(
        [Ig2_Sphere_Sphere_ScGeom6D()],
        [Ip2_CohFrictMat_CohFrictMat_CohFrictPhys(
            setCohesionNow=True, setCohesionOnNewContacts=False)],
        [Law2_ScGeom6D_CohFrictPhys_CohesionMoment()]
    ),
    NewtonIntegrator(gravity=(0, -g, 0), damping=damp, label='newton'),
]

# define material:
O.materials.append(CohFrictMat(young=1e5, poisson=0, density=1e1,
                               frictionAngle=math.radiansHP1(0), normalCohesion=1e7,
                               shearCohesion=1e7, momentRotationLaw=False, label='mat'))

# create spheres
nodeIds = []
for i in range(0, n):
    nodeIds.append(O.bodies.append(sphere([i*L/n*math.cos(inclination),
        i*L/n*math.sin(inclination), 0], r, wire=False, fixed=False,
        material='mat', color=color)))

# create rods
for i, j in zip(nodeIds[:-1], nodeIds[1:]):
    inter = createInteraction(i, j)
    inter.phys.unp = -(O.bodies[j].state.pos-O.bodies[i].state.pos).norm() + \
        O.bodies[i].shape.radius+O.bodies[j].shape.radius

O.bodies[0].dynamic = False  # set a fixed upper node
qt.View()                    # create a GUI view
Gl1_Sphere.stripes = True    # mark spheres with stripes
rr = qt.Renderer()           # get instance of the renderer
rr.intrAllWire = True        # draw wires
rr.intrPhys = True           # draw the normal forces between the spheres.
\end{lstlisting}

\end{document}